\documentclass[useAMS,usenatbib]{mn2e}
\usepackage{amsmath, amssymb}
\usepackage{graphicx}
\usepackage{wasysym}
\usepackage{color}
\usepackage{longtable}

\newcommand{\eg}[0]{$\textnormal{e.g. }$}
\newcommand{\ie}[0]{$\textnormal{i.e. }$}
\newcommand{\Msun}[0]{\,\textnormal{M}_{\textnormal{\astrosun}}}
\newcommand{\asun}[0]{_{\textnormal{\astrosun}}}

\newcommand{\tn}[1]{\textnormal{#1}}
\newcommand{\sub}[1]{_{\textnormal{#1}}}
\newcommand{\bs}[1]{\boldsymbol{#1}}

\newcommand{\Temp}[0]{\tn{log}(kT_{500}/\tn{keV})}
\newcommand{\NumObsSystems}[0]{79 }
\newcommand{\NumObsSystemsN}[0]{79}
\newcommand{\NumObsMeasurements}[0]{159 }
\newcommand{\ObsTZslope}[0]{-0.26 }

\newcommand{\ObsTZslopeErr}[0]{0.03 }
\newcommand{\ObsTZnorm}[0]{0.37 } 
\newcommand{\ObsTZnormErr}[0]{0.02 }
\newcommand{\ObsTZscatter}[0]{0.10 }

\newcommand{\ObsTZslopeGroups}[0]{0.73 }
\newcommand{\ObsTZslopeGroupsErr}[0]{0.17 }
\newcommand{\ObsTZnormGroups}[0]{0.61 } 
\newcommand{\ObsTZnormGroupsErr}[0]{0.03 }
\newcommand{\ObsTZscatterGroups}[0]{0.17 }
\newcommand{\ModelTZslope}[0]{-0.10 }

\title[Iron in groups \& clusters]{Iron in galaxy groups and clusters: Confronting galaxy evolution models with a newly homogenised dataset}
\author[Yates et al.]{Robert M. Yates$^{1,2}$\thanks{Email: robyates@mpe.mpg.de}, Peter A. Thomas$^{3}$ \& Bruno M. B. Henriques$^{2,4}$\\\\
$^{1}$ Max Planck Institut f$\ddot{u}$r Extraterrestrische Physik, Giessenbachstra\ss{}e 1, 85748, Garching, Germany\\
$^{2}$ Max Planck Institut f$\ddot{u}$r Astrophysik, Karl-Schwarzschild-Str. 1, 85741, Garching, Germany\\
$^{3}$ Astronomy Centre, University of Sussex, Falmer, Brighton BN1 9QH, UK\\
$^{4}$ Institute for Astronomy, Department of Physics, ETH Zurich, 8093 Zurich, Switzerland}

\begin{document}
\date{Accepted ??. Received ??; in original form ??}
\maketitle

\begin{abstract}
We present an analysis of the iron abundance in the hot gas surrounding galaxy groups and clusters. To do this, we first compile and homogenise a large dataset of \NumObsSystems low-redshift ($\left|z\right| = 0.03$) systems (\NumObsMeasurements individual measurements) from the literature. Our analysis accounts for differences in aperture size, solar abundance, and cosmology, and scales all measurements using customised radial profiles for the temperature ($T$), gas density ($\rho\sub{gas}$), and iron abundance ($Z\sub{Fe}$). We then compare this dataset to groups and clusters in the \textsc{L-Galaxies} galaxy evolution model.

Our homogenised dataset reveals a tight $T$-$Z\sub{Fe}$ relation for clusters, with a scatter in $Z\sub{Fe}$ of only $\ObsTZscatter$ dex and a slight negative gradient. After examining potential measurement biases, we conclude that at least some of this negative gradient has a physical origin. Our model suggests greater accretion of hydrogen in the hottest systems, via stripping of gas from infalling satellites, as a cause. At lower temperatures, \textsc{L-Galaxies} over-estimates $Z\sub{Fe}$ in groups, indicating that metal-rich gas removal (via \eg AGN feedback) is required.

\textsc{L-Galaxies} provides a reasonable match to the observed $Z\sub{Fe}$ in the intracluster medium (ICM) of the hottest clusters from at least $z\sim 1.3$ to 0.3. This is achieved without needing to modify any of the galactic chemical evolution (GCE) model parameters. However, the $Z\sub{Fe}$ in intermediate-$T$ clusters appears to be under-estimated in our model at $z=0$. The merits and problems with modifying the GCE modelling to correct this are discussed.
\end{abstract}


\section{Introduction} \label{sec:Introduction}
\LARGE{G}\normalsize alaxy clusters have long been known to host large reservoirs of hot gas \citep{M76}. This intracluster medium (ICM) is a mix of accreted pristine gas and enriched material that has been processed through stars and driven out of the member galaxies via supernov\ae{}, stellar winds, stripping processes, and feedback from active galactic nuclei (AGN). The same is true of galaxy groups, which contain less hot gas than clusters but are much more numerous. Studying the chemical evolution of the ICM therefore provides distinct insights into three of the most fundamental questions in galaxy evolution -- what comes in, what goes out, and when does this infall and outflow occur?

Observational studies of emission lines in the X-ray spectra of nearby clusters have found that the local ICM is enriched with iron to around one third of the solar abundance (\eg \citealt{ES91,F98,DGM01,Ta04}). There is also some indication that this enrichment was largely complete by $z\sim1$ (\eg \citealt{ML97,AF98,To03,An09,Ba12}), and similar conclusions can be drawn from observations of the cool circumgalactic medium (CGM) surrounding massive galaxies at $z\sim2$ \citep{P14}.

The large amount of metals detected in the ICM of nearby clusters has, however, posed a long-standing problem for galaxy evolution models. Super-solar iron yields were required by early models in order to reproduce the observed ratio between iron mass in the ICM and B-band luminosity of the cluster galaxies \citep{KC98,DL04}. More recently, the purely analytic model presented by \citet{RA14} also suggests that the iron yield in the largest clusters ($M_{500} > 10^{14}\Msun$) needs to be four times the solar value. And even studies which have incorporated more sophisticated modelling of the production and distribution of chemical elements have found that, when assuming typical stellar yields, the ICM iron abundance is around 0.25 dex below that observed \citep{N05a,A10b}.

From the model perspective, possible solutions to this problem have included changes to the shape of the stellar initial mass function (IMF), increases in the efficiency of iron production in SNe-Ia, more efficient metal ejection from galaxies, and including pair-instability supernov\ae{} (\eg \citealt{Mo03,N05a,Bo08,Fa10,A10b,Sh13,Mo14}). However, such changes also have a significant impact on the chemical compositions of the galaxies within clusters and in lower-density environments (\S \ref{sec:Iron_in_model_clusters} and \citealt{A10b}).

In this work, our interest in ICM enrichment is twofold: Firstly, we wish to obtain a large, homogenised dataset of iron abundance ($Z\sub{Fe}$) measurements for local groups and clusters. The correlation between $Z\sub{Fe}$ and the ICM temperature ($T$) can then be studied in unprecedented detail. Secondly, we will use this dataset to test the Munich semi-analytic model of galaxy evolution, \textsc{L-Galaxies}. We wish to determine if our model can reproduce the key trends observed in the ICM without compromising its good agreement with the chemical composition of other astrophysical regions.

This paper is structured as follows: In \S \ref{sec:Obs_sample}, we outline our observational dataset, the classifications we adopt and our methods for homogenising measurements of ICM temperature and iron abundance. The ten observational samples we utilise are discussed individually in Appendix A. In \S \ref{sec:Obs_results}, we show the $T$-$Z\sub{Fe}$ relation for our homogenised dataset of nearby groups and clusters, and discuss the revealed trends in detail. In \S \ref{sec:The_semi-analytic_model}, we describe the core version of the \textsc{L-Galaxies} galaxy evolution model. In \S \ref{sec:Model sample}, we outline our model sample of groups and clusters, and present how their properties are scaled to match those observations to which we compare. In \S \ref{sec:Model results}, we discuss how \textsc{L-Galaxies} performs when compared to observations of the baryon fraction in clusters, the $T$-$Z\sub{Fe}$ relation and the evolution of $Z\sub{Fe}$ with redshift. In \S \ref{sec:Conclusions}, we provide a summary of our results and our conclusions. Where the model is able to reproduce the data, we investigate the physical mechanisms modelled that have lead to these results. Where the model fails, we discuss possible ways to improve the agreement. 

Throughout this work, the logarithm of $x$ to the base ten is written simply as log$(x)$.

\section{Observational sample} \label{sec:Obs_sample}
Here, we outline how systems from our observational dataset are classified, and how key properties are calculated. The 10 observational samples that we consider, along with the acronyms we adopt for them hereafter, are listed in Table \ref{tab:Obs_samples}.

Different methods have been required when processing different samples, depending on the specifics of the survey from which they were obtained. However, the following classifications are always applied where possible. Any departures from this set of definitions for specific systems are detailed in Appendix A.

\begin{table*}
\centering
\begin{tabular}{lccccccc}
\hline \hline
\textbf{Study}  & \textbf{Acronym} & \textbf{Observatory} & \multicolumn{5}{c}{\textbf{Systems}} \\
 & & & Groups & Clusters & NCC & CC & Total \\
\hline
\citet{F98} & F98 & \textit{ASCA} & 6 & 28 & 9 & 25 & \textbf{34} \\
\citet{DGM01} & DGM01 & \textit{BeppoSAX} & - & 17 & 8 & 9 & \textbf{17} \\
\citet{P03} & P03 & \textit{XMM Newton} & 1 & 10 & - & 11 & \textbf{11} \\
\citet{Ta04} & T04 & \textit{XMM Newton} & 1 & 16 & 4 & 13 & \textbf{17} \\
\citet{dP07} & dP07 & \textit{XMM Newton} & - & 21 & 4 & 17 & \textbf{21} \\
\citet{M11} & M11 & \textit{XMM Newton} & - & 26 & 6 & 20 & \textbf{26} \\
 & & & & & & & \\
\citet{Mah05} & M05 & \textit{XMM Newton} & 7 & 1 & 1 & 7 & \textbf{8} \\
\citet{Fi06} & F06 & \textit{XMM Newton} & 6 & - & 1 & 5 & \textbf{6} \\
\citet{RP09} & RP09 & \textit{Chandra} & 14 & 1 & 1 & 14 & \textbf{15} \\
\citet{S14} & S14 & \textit{Suzaku} & 4 & - & - & 4 & \textbf{4} \\
\hline
 & & & 39 (25) & 120 (54) & 34 (21) & 125 (58) & \textbf{\NumObsMeasurements (\NumObsSystemsN)} \\ 
\hline \hline
\end{tabular}
\caption{The observational samples we consider in our dataset. The number of usable $T$ and $Z\sub{Fe}$ measurements from each sample for groups \& clusters and NCC \& CC are shown. Counts in parenthesis give the total number of \textit{unique} systems considered.}
\label{tab:Obs_samples}
\end{table*}

\subsection{Definitions} \label{sec:Definitions_and_Derivations}
\begin{itemize}
\item \textit{Groups and clusters}: We choose to distinguish between galaxy groups and clusters by their ICM temperature, with a threshold value of $\Temp = 0.1$ (or $kT_{500} = 1.26$ keV), where $T_{500}$ is the temperature \textit{at} $r_{500}$. This corresponds to a mean, emission-weighted, ICM temperature of $k\bar{T}\sub{500,ew} \sim 1.9$ keV. We acknowledge that this is an overly simplistic definition, but note that our choice is similar to the value of $k\bar{T}\sub{500,ew}=2.0$ keV chosen by previous authors for their studies of groups and clusters (\eg \citealt{M03} and \citealt{Mah05}).

\item \textit{Cool and non-cool cores}: We further classify all objects as exhibiting either a cool-core (CC) or a non-cool-core (NCC). It should be noted that the groups and clusters in our dataset actually form a continuous distribution of systems, with mass deposition rates ($\dot{M}\sub{dep}$) ranging from 0 to $\sim 1000 \Msun/\tn{yr}$ (see \eg \citealt{Pe98}, fig. 7) and temperature gradients ranging from steeply negative to positive. Nonetheless, some distinction between CC and NCC systems is required, as they are known to have distinct properties which influence our estimation of the iron abundance.

Therefore, we match clusters to the \citeauthor{H10} (2012, table 3) catalogue of CC and NCC systems where possible. Their classification is based on measurements of the central cooling time (CCT), where clusters with $\tn{CCT} > 7.59\,h_{73}^{-1/2}\,\tn{Gyr}$ are defined as being NCC. For the 17 of our clusters not in the \citet{H10} sample, we rely on the $\dot{M}\sub{dep}$ measured by \citet{Wh97} or \citet{Pe98}, who utilised \textit{Einstein} and \textit{ROSAT} data. In these cases, all objects with $\dot{M}\sub{dep}$ consistent with zero within errors are classified as NCC.

Under this classification scheme, 33 per cent of our sample of 54 clusters are NCCs. This is slightly higher than the 28 per cent found by \citet{H10} for their sample. This difference could be due to our reliance on the less-accurate $\dot{M}\sub{dep}$ determination for some objects. The majority (59 per cent) of the 17 clusters \textit{not} present in the \citet{H10} sample are designated as NCCs according to their $\dot{M}\sub{dep}$, which can be under-predicted when using \textit{Einstein} and \textit{ROSAT} data.

We note that only five of our clusters have discrepant classifications when considering $\dot{M}\sub{dep}$ or CCT as the indicator. All five are classified as NCC when using $\dot{M}\sub{dep}$ and as weak CCs when using CCT. With the exception of A2589, all these clusters are known to be disturbed (\ie merging), so this discrepancy is likely due to these systems losing their cool cores and becoming NCCs.

For galaxy groups, we largely rely on either the temperature profiles measured by \citet{RP07}, or the classifications of \citet{J09}, who defined CC systems as those with a mean temperature within $0.1-0.3\,r_{500}$ greater than that within $0.00-0.05\,r_{500}$.

Groups or clusters that are not present in any of the above-mentioned catalogues were determined to be CC or NCC based on other temperature profile measurements from the literature (see Appendix A).

\item \textit{Hubble parameter:} Where necessary, physical properties were corrected for differences in the assumed dimensionless Hubble parameter, $\textnormal{h}\equiv\tn{H}_{0}/(100$ km s$^{-1}\,$Mpc$^{-1})$, rescaling to $h = 0.73$ to match that assumed in our galaxy formation model (\S \ref{sec:The_semi-analytic_model}). We note that $h$ is factored-in to all numbered quantities where necessary in this work, with the placeholder $h_{73}=1$ included to indicate our assumed cosmology.

\item \textit{Structural parameters}: For clusters, values of redshift ($z$), $r_{200}$, and $M_{200}$ (required to obtain $r_{500}$), as well as $r\sub{c}$ and $\beta$ (required to calculate the gas density profile), were exclusively taken from the catalogue of \citet{RB02}, to ensure that they are calculated in a consistent way. This means that six objects from our cluster samples are not in our final data set due to not being present in the \citet{RB02} extended catalogue. For groups, $z$, $r\sub{c}$ and $\beta$ were taken from the \citet{M03} or \citet{RP09} group catalogues where possible, and $r_{500}$ was estimated from the mean temperature, as described in \S \ref{sec:Radius estimation}. For those groups where structural parameters could not be obtained from the literature, we estimated $r\sub{c}$ and $\beta$ in the same way as for our model sample (\S \ref{sec:Model sample}).

\item \textit{Solar abundances}: All chemical abundances were re-normalised to the solar photospheric abundances provided by \citet{GS98}. On this scale, the solar abundance of iron by number is $N\sub{Fe,\astrosun}/N\sub{H,\astrosun}=3.16\times10^{-5}$. This is an important step in the homogenisation process, as, for example, there is a 0.17 dex drop in the estimated iron abundance when using the solar photospheric value of $N\sub{Fe,\astrosun}/N\sub{H,\astrosun}=4.67\times10^{-5}$ measured by \citet{AG89} rather than that of \citet{GS98}.
\end{itemize}

\subsection{Radius estimation} \label{sec:Radius estimation}
We choose to scale all measurements to $r_{500}$, the radius within which the matter density is 500 times that of the critical density of the Universe (\ie enclosing an over-density of $\Delta_{500}\equiv500$). This radius is chosen to minimise the degree of correction required for observational measurements, which are typically taken between $r_{2500}$ and $r_{500}$.

For galaxy clusters, to ensure consistency with the way this radius is calculated in our galaxy evolution model, $r_{500}$ is obtained from published values of $r_{200}$ and $M_{200}$ by

\begin{equation}
r_{500} = a\,x_{500}\;\;,
\end{equation}
where $a=r_{200}/c$ is the scale length, and the concentration, $c$, is assumed to be given by

\begin{equation}
c = \left(\frac{9.90}{1+z}\right)\left(\frac{M_{200}}{1\times10^{4}\,h_{73}^{-1}}\right)^{-0.102}\;\;,
\end{equation}
as found by \citet{Dol04} for simulated haloes in a $\Lambda$CDM cosmology with $\sigma_{8}=0.9$.

The quantity $x_{500}$ is obtained assuming an NFW dark matter (DM) density profile \citep{NFW97}, given by

\begin{equation}
\rho\sub{DM}(r)=\frac{\rho_{0}}{(r/a)(1+r/a)^{2}}\;\;,
\end{equation}
where $\rho_{0} = \rho\sub{crit}\,\delta\sub{c} = (3H_{0}^{2}/8\pi G)\delta\sub{c}$, and $\delta\sub{c}$ is the characteristic density, given by

\begin{equation}
\delta\sub{c} = \frac{\Delta_{200}}{3}\frac{c^{3}}{\tn{ln}(1+c)-c/(1+c)}\;\;.
\end{equation}

We note that using the values of $r_{500}$ provided by \eg \citet{RB02}, rather than values rescaled from $r_{200}$, makes a negligible difference to our results.

For galaxy groups, measured values of $r_{200}$ and $M_{200}$ are very rare in the literature. We therefore estimate $r_{500}$ from the mean, emission-weighted temperature, following the scaling relation presented by \citet{Fi01b};

\begin{equation}\label{eqn:r-T_relation_RP07}
r_{500} = 0.432\tn{\scriptsize{$\pm0.007$}}\,\sqrt{k\bar{T}\sub{ew}}\,h_{73}^{-1}\,\tn{Mpc}\;\;.
\end{equation}

\begin{figure}
\centering
\includegraphics[width=0.45\textwidth]{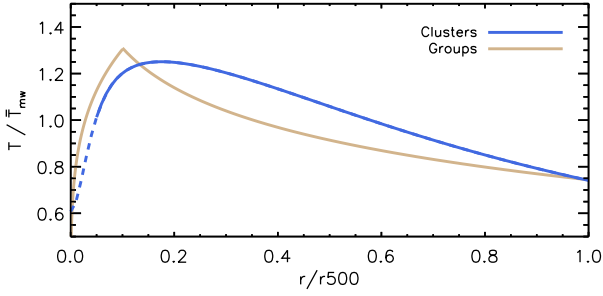} \\
\caption{The default temperature profiles assumed for clusters and groups, taken from \citet{V06} and \citet{RP07}, respectively (\S \ref{sec:Temperature estimation}). The mean temperature for the group profile has been re-scaled in this figure to a mass-weighted value using the spectroscopic-to-mass weighted conversion factor of 1.11 provided by \citet{V06}. The dashed line indicates an extrapolation beyond the radii in which the \citet{V06} cluster profile was fit.}
\label{fig:default_T_profiles}
\end{figure}

\subsection{Temperature estimation} \label{sec:Temperature estimation}
The X-ray-emitting-gas temperature \textit{at} $r_{500}$ is obtained using distinct temperature profiles for groups and clusters.

For clusters, $T_{500}$ is obtained using the typical temperature profile found by \citet{V06} for their systems with $kT > 2.5$ keV:

\begin{equation}\label{eqn:temp_profile_V06}
T(r) = 1.35\,\bar{T}\sub{mw}\left[\frac{(x/0.045)^{1.9}+0.45}{(x/0.045)^{1.9}+1}\right]\left[\frac{1}{1+(x/0.6)^{2}}\right]\;\;,
\end{equation}
where $x=r/r_{500}$ and $\bar{T}\sub{mw}$ is the mean, mass-weighted gas temperature. Where necessary, measured peak or emission-weighted temperatures are converted to mass-weighted values using the following conversion factors,

\begin{equation}\label{eqn:temp_conversions_V06}
T\sub{peak}\,:\,\bar{T}\sub{ew}\,:\,\bar{T}\sub{mw} = 1.21\,:\,1.11\,:\,1\;\;,
\end{equation}
provided by \citeauthor{V06} (2006, eqn. 9).

For groups, the preferred temperature profile of \citeauthor{RP07} (2007, eqns. 3 and 7) for their sample of galaxy groups is used:

\begin{align}\label{eqn:temp_profile_RP07}
\nonumber T(r) &= \bar{T}\sub{ew}\cdot{}x^{0.21\pm 0.02}\, 10^{0.28\pm 0.03} & \tn{for }x<0.1 & \\
 &= \bar{T}\sub{ew}\left[-0.51\tn{\scriptsize{$\pm 0.04$}}\,\tn{log}(x) + 0.67\tn{\scriptsize{$\pm 0.03$}}\right]& \tn{for }x\geq0.1 & \;\;,
\end{align}
where $x=r/r_{500}$.

These two profiles are shown in Fig. \ref{fig:default_T_profiles}. We can see that, when normalised by mean mass-weighted temperature, both provide essentially identical values of $T_{500}\sim 0.74\,\bar{T}\sub{mw}$.\footnote{We note that Eqn. \ref{eqn:temp_profile_V06} for clusters was fit using mean temperatures measured between $r=69\,h_{73}^{-1}$kpc  ($\sim 0.005$ - $0.01\,r_{500}$) and $r_{500}$, whereas Eqn. \ref{eqn:temp_profile_RP07} relied on mean temperatures measured between $0.1$ and $0.3\,r_{500}$.} Indeed, when comparing the two temperature profiles using our dataset, we find the value of $T_{500}$ obtained only differs by $\sim 0.02$ keV or less [as low as $\sim 0.005$ keV for objects with temperatures around the transition value of $\Temp=0.1$]. \citet{RP07} also found a similarity between group and cluster temperature profiles when comparing various forms, including Eqn. \ref{eqn:temp_profile_RP07} and a \citet{V06} profile analogue.

We do not adopt different temperature profiles for CC and NCC systems, even though these two classes self-evidently have different core temperatures, as their temperature profiles beyond the core (\eg at around $r_{500}$) are found to be quite similar (\eg \citealt{M98,DGM02,LM08}). The mean temperatures quoted should be representative of the average ICM temperature beyond the core, and so can be scaled appropriately for both CC and NCC systems with our cluster profile.

\begin{figure}
\centering
\includegraphics[width=0.45\textwidth]{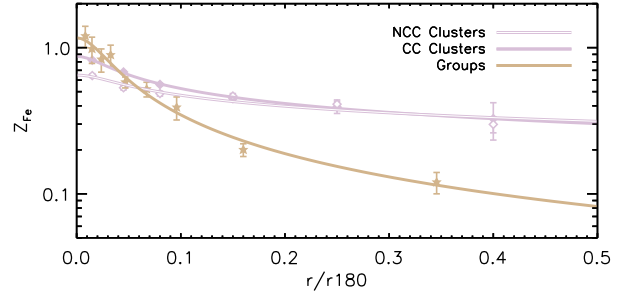} \\
\caption{The default iron abundance profiles assumed for NCC clusters, CC clusters, and groups (\S \ref{sec:Iron abundance estimation}). The group profile has been rescaled to $r_{180}$ in this figure, assuming $r_{500} = 0.64\,r_{180}$.}
\label{fig:default_Z_profiles}
\end{figure}

\subsection{Iron abundance estimation} \label{sec:Iron abundance estimation}
Iron abundances in this work are given as

\begin{equation}\label{eqn:Abundances}
\textnormal{log}(Z\sub{Fe}) = \textnormal{log}\left(\frac{N\sub{Fe}}{N\sub{H}}\right) - \textnormal{log}\left(\frac{N\sub{Fe,\astrosun}}{N\sub{H,\astrosun}}\right)\;\;,
\end{equation}
where $N\sub{Fe}$ is the number of iron atoms, and $N\sub{Fe,\astrosun}/N\sub{H,\astrosun}$ is the iron abundance by number in the solar photosphere, assumed here to be given by \citet{GS98}.

In order to obtain the mass-weighted iron abundance within $r_{500}$ ($\bar{Z}\sub{Fe,500}$) from measurements of the emission-weighted iron abundance within some observed aperture ($\bar{Z}\sub{Fe,obs}$), assumptions need to be made about the distribution of gas and iron within the ICM. For the gas density profile, we assume a single $\beta$ model \citep{CFF76}, given by

\begin{equation}\label{eqn:rho_profile}
	\rho_{\tn{gas}}(r) = \rho\sub{gas,0}\left[1+\left(\frac{r}{r\sub{c}}\right)^{2} \right]^{-3\beta/2}\;\;,
\end{equation}
with the slope, $\beta$, and core radius, $r\sub{c}$, obtained from the literature for each object individually. Although this single $\beta$ profile is often assumed in theoretical studies of model clusters (\eg \citealt{N05a}; \citealt{A10b}), we caution that it is not necessarily accurate for all systems in reality (see \eg \citealt{M03,V06}), and is therefore only an approximation.

We also assume a $\beta$ model for the iron abundance profile, as chosen by \citet{DG04} for their cluster sample, given by
\begin{equation}\label{eqn:Z_profile_DGM01}
	Z\sub{Fe}(r) = Z\sub{Fe,0}\left[1+\left(\frac{x}{x_{\tn{c}}}\right)^{2} \right]^{-\alpha}\;\;,
\end{equation}
where $x = r/r_{180}$ for clusters and $r/r_{500}$ for groups. However, we re-calculate the parameters $x_{\tn{c}}$ and $\alpha$ to fit more recent data. For clusters, we fit Eqn. \ref{eqn:Z_profile_DGM01} to the binned radial abundance data provided by M11 (their table 4) for 16 CC clusters, and also separately for 10 NCC clusters. For groups, we fit Eqn. \ref{eqn:Z_profile_DGM01} to the binned radial abundance data provided by \citet{RP07} for 15 galaxy groups. The fitting parameters we obtain are given in Table \ref{tab:Z_profile_params}, and the three default iron abundance profiles are shown in Fig. \ref{fig:default_Z_profiles}.

\begin{table}
\centering
\begin{tabular}{lcccc}
\hline \hline
\textbf{ } & $\bs{x_{\tn{\textbf{c}}}}$ & $\bs{\sigma(x_{\tn{\textbf{c}}})}$ & $\bs{\alpha}$ & $\bs{\sigma(\alpha)}$ \\
\hline
NCC clusters & 0.031 & 0.028 & 0.132 & 0.044 \\
CC clusters & 0.025 & 0.003 & 0.176 & 0.006 \\
Groups & 0.027 & 0.001 & 0.322 & 0.005 \\
\hline \hline
\end{tabular}
\caption{The fitting parameters, and their errors, for the three default iron abundance profiles we assume in this work (plotted in Fig. \ref{fig:default_Z_profiles}). These profiles are fit to data from M11 for the CC and NCC cluster profiles, and data from \citet{RP07} for the group profile. We note that $x\sub{c}$ is the characteristic radius normalised to $r_{180}$ for the cluster profiles and to $r_{500}$ for the group profile.}
\label{tab:Z_profile_params}
\end{table}

Unlike the temperature profiles described in \S \ref{sec:Temperature estimation}, these different iron abundance profiles lead to significantly different estimates of the mean iron abundance. For example, for any given object, our CC cluster profile returns a value of $\bar{Z}\sub{Fe,500}$ which is around 0.10 dex higher than our group profile. This is because the gradient for groups is steeper than for clusters, in line with the findings of \citet{RP07}. More of the total iron is concentrated close to the centre in groups, causing the correction factor when calculating the mean iron abundance within $r_{500}$ to be larger.

We consider the choice of iron abundance profile to be one of the major sources of uncertainty in our results. In practice, individual systems can have iron abundance profiles that are quite distinct from any typical profile, even within the samples from which these typical profiles are determined.

Given these default profiles, we obtain $\bar{Z}\sub{Fe,500}$ following a two step process: Firstly, we normalise the iron abundance profile for each system, by obtaining $Z\sub{Fe,0}$ from the observed \textit{emission-weighted} iron abundance (measured within $r\sub{min}<r<r\sub{max}$):

\begin{equation} \label{eqn:scaled_Z_lw}
Z\sub{Fe,0} = \bar{Z}\sub{Fe,obs}\,\frac{\int_{r\sub{min}}^{r\sub{max}}\rho^{2}_{\tn{gas}}\,r^{2}\,\tn{dr}}{\int_{r\sub{min}}^{r\sub{max}}\left[1+(x/x\sub{c})^{2}\right]^{-\alpha}\,\rho^{2}_{\tn{gas}}\,r^{2}\,\tn{dr}}\;\;.
\end{equation}

We then use this normalised profile to obtain the \textit{mass-weighted} iron abundance within $r_{500}$, given by

\begin{equation}\label{eqn:scaled_Z}
	\bar{Z}\sub{Fe,500} = \frac{\int_{0}^{r_{500}}Z\sub{Fe}\,\rho_{\tn{gas}}\,r^{2}\,\tn{dr}}{\int_{0}^{r_{500}}\rho_{\tn{gas}}\,r^{2}\,\tn{dr}}\;\;.
\end{equation}

This procedure allows us to obtain scaled, mass-weighted iron abundances for any galaxy group or cluster with an X-ray spectrum, even if no radial information is available. In this work, we will show how this procedure yields $\bar{Z}\sub{Fe,500}$ estimates that are consistent among different studies of the same system (\S \ref{sec:An object-for-object comparison}), and with alternative estimates of the mass-weighted iron abundance using the same data (\S A.2.3).

\begin{figure*}
\centering
\includegraphics[width=0.95\textwidth]{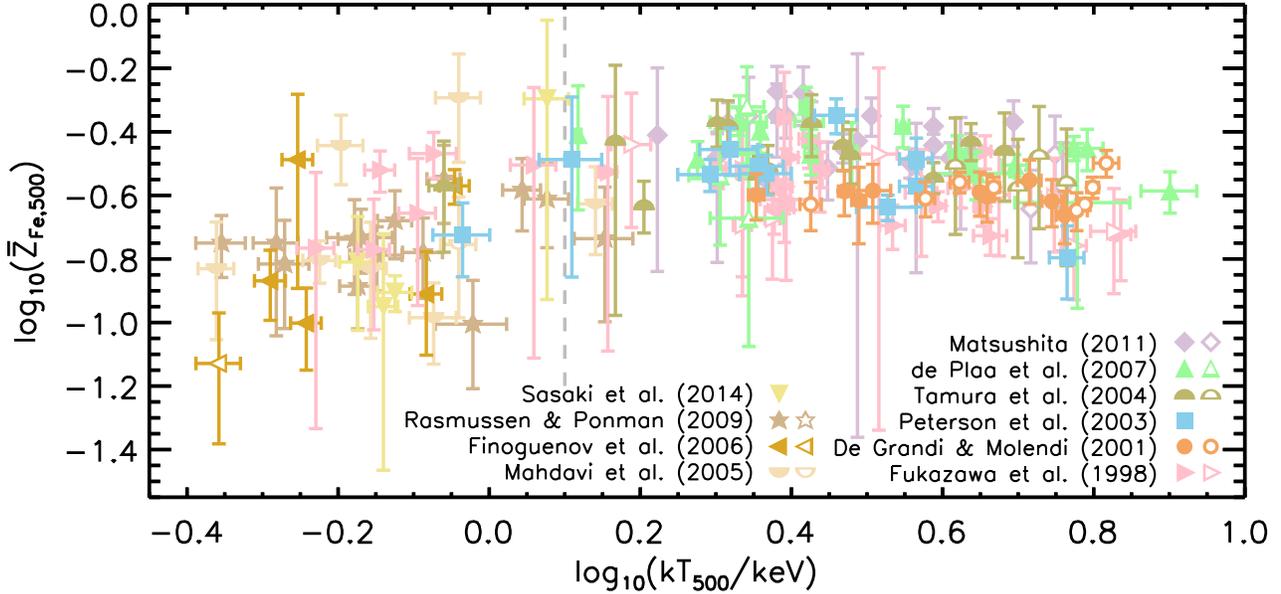}
\caption{The $kT_{500}$-$\bar{Z}\sub{Fe,500}$ relation for our whole dataset of observed local groups and clusters. Filled symbols indicate CC systems and open symbols indicate NCC systems. The grey vertical line separates groups and clusters at $\Temp = 0.1$, as discussed in \S \ref{sec:Definitions_and_Derivations}. There is a strong $T$-$Z\sub{Fe}$ correlation for groups, albeit with a large scatter, and a weak $T$-$Z\sub{Fe}$ anti-correlation for clusters, with a small scatter of only \ObsTZscatter dex.}
\label{fig:kT-FeH_Obs}
\end{figure*}

\section{Observational results}\label{sec:Obs_results}
Fig. \ref{fig:kT-FeH_Obs} shows the $kT_{500}$ - $\bar{Z}\sub{Fe,500}$ relation for our complete dataset of galaxy groups and clusters. We can see that clusters tend to have higher iron abundances than groups. Below $\Temp\sim 0.2$, there is a systematic decrease in $Z\sub{Fe}$ with decreasing temperature. This trend is discussed in \S \ref{sec:T-Z for groups}. Above $\Temp\sim 0.3$, the $T$-$Z\sub{Fe}$ relation for clusters is remarkably tight and has a weak negative slope. This trend is discussed in \S \ref{sec:T-Z for clusters}.

Errors on all measurements and fitting parameters have been fully propagated through to a final error in $\bar{Z}\sub{Fe,500}$ in this work. These final errors are predominantly driven by the uncertainty in the original iron abundance measurement, with the exception of a few systems whose core radius, $r\sub{c}$, is poorly constrained by the measured surface brightness profile (\eg A2147). These systems have larger errors on $\bar{Z}\sub{Fe,500}$, as the metallicity determination is particularly sensitive to any uncertainty in $r\sub{c}$.

The data plotted in Fig. \ref{fig:kT-FeH_Obs} is provided in Tables \ref{tab:Final_data0}, \ref{tab:Final_data1}, and \ref{tab:Final_data2} at the end of this work, and is available online.\footnote{Our full observational dataset is freely available online in electronic format at \textit{robyatesastro.moonfruit.com/data/}}

\begin{figure*}
\centering
\includegraphics[width=0.95\textwidth]{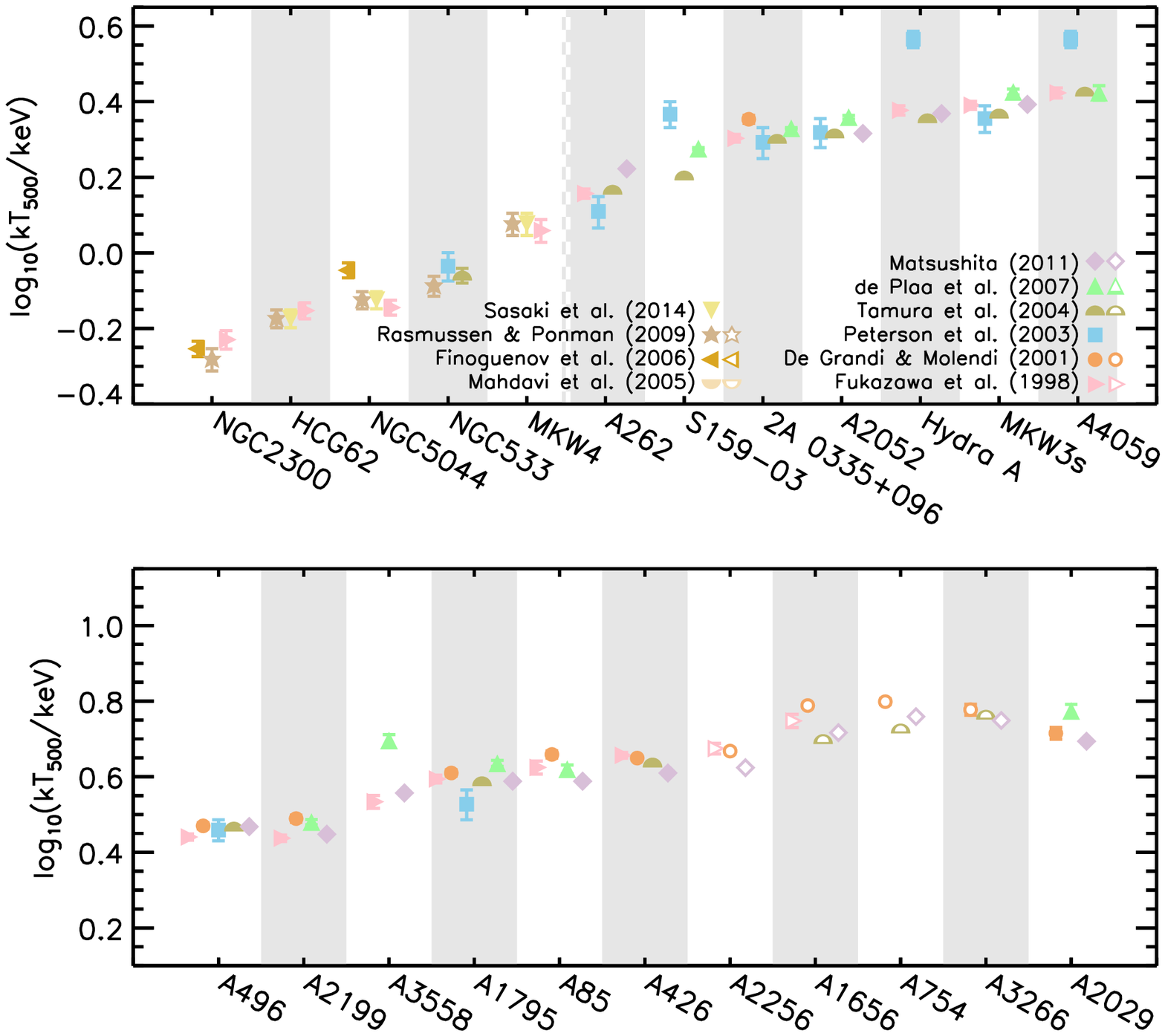}
\caption{A comparison between the values of $kT_{500}$ derived for the same systems from different studies. Only systems with three or more separate estimates of $T_{500}$ and $\bar{Z}\sub{Fe,500}$ are shown. Objects are ordered from left to right by ICM temperature. The dashed vertical line in the top panel separates groups from clusters, as discussed in \S \ref{sec:Definitions_and_Derivations}.}
\label{fig:Temp_comp_Obs}
\end{figure*}

\begin{figure*}
\centering
\includegraphics[width=0.95\textwidth]{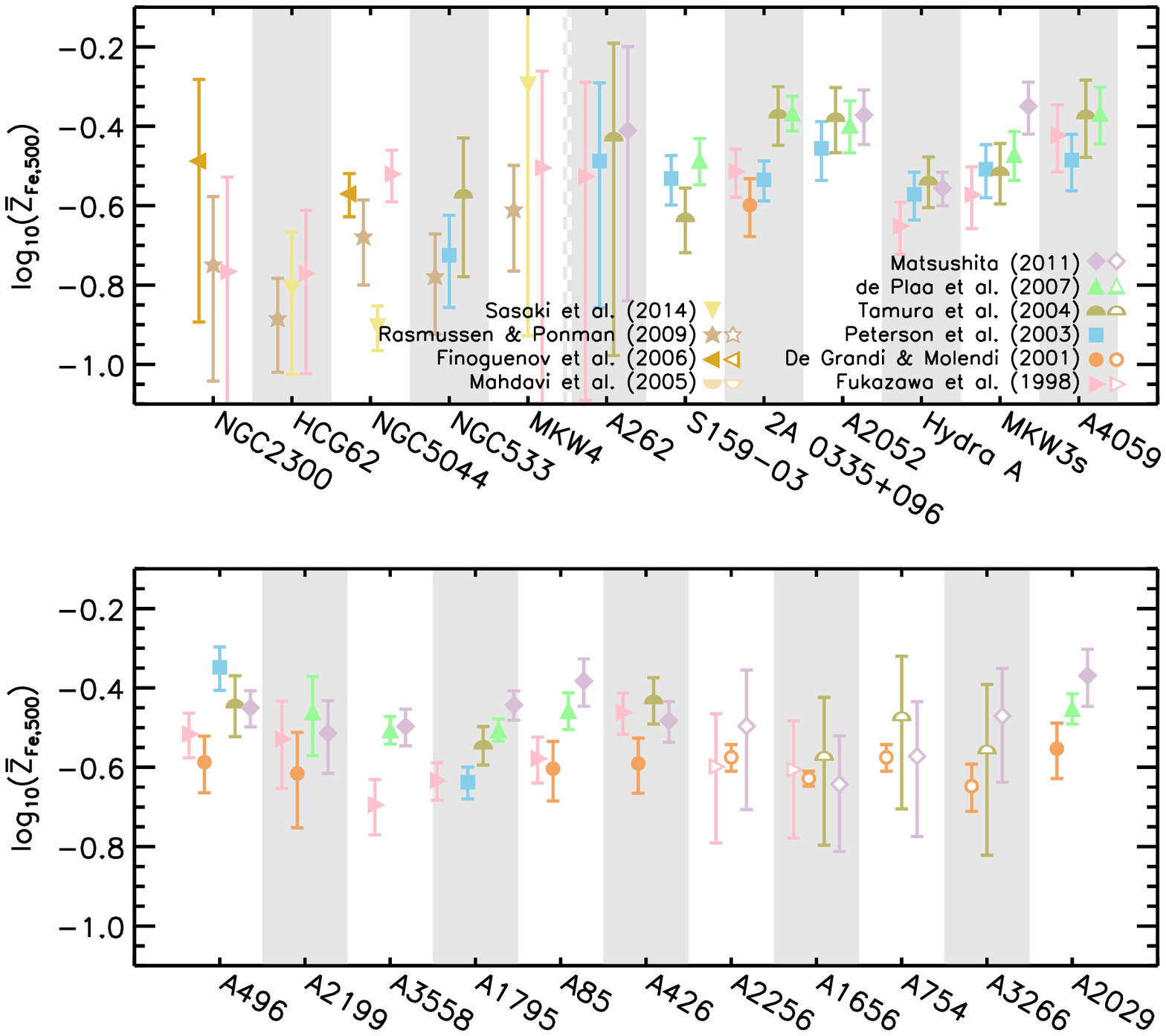}
\caption{A comparison between the values of $\bar{Z}\sub{Fe,500}$ derived for the same systems from different studies. As in Fig. \ref{fig:Temp_comp_Obs}, only systems with three or more separate estimates of $T_{500}$ and $\bar{Z}\sub{Fe,500}$ are shown. Objects are ordered from left to right by ICM temperature. The dashed vertical line in the top panel separates groups from clusters, as discussed in \S \ref{sec:Definitions_and_Derivations}.}
\label{fig:FeH_comp_Obs}
\end{figure*}

\subsection{The $T$-$Z\sub{Fe}$ relation for clusters}\label{sec:T-Z for clusters}
A linear fit to the data above $\Temp=0.25$ yields the following relation for clusters:

\begin{equation}\label{eqn:TZR for clusters}
	\tn{log}(\bar{Z}\sub{Fe,500}) = (\ObsTZslope\tn{\scriptsize{$\pm \ObsTZslopeErr$}})\,\Temp - (\ObsTZnorm\tn{\scriptsize{$\pm \ObsTZnormErr$}})
\end{equation}
with a $1\sigma$ dispersion in $\bar{Z}\sub{Fe,500}$ from residuals of \mbox{\ObsTZscatter dex.} The $T$-$Z\sub{Fe}$ relation for galaxy clusters is therefore as tight as the well-established $M_{*}$-$Z\sub{O}$ relation (\ie the mass-metallicity relation) for local, star-forming galaxies \citep{T04,Y13}.

The slight decrease in $\bar{Z}\sub{Fe,500}$ with increasing temperature apparent in Fig. \ref{fig:kT-FeH_Obs} is also seen in most of the individual cluster (\S A.1.7). Such an anti-correlation has already been observed and discussed in the literature (\eg F98; P03; \citealt{Ba05,Ba07}; M11; \citealt{Ho16}), and the possibility that this is an artificial effect must be assessed.

F98 suggested that a higher central concentration of iron in lower-temperature clusters could bias abundance measurements when observing the cores. If this were the case, then our choice of a fixed slope for the $Z\sub{Fe}$ profile would be artificially increasing the estimated value of  $\bar{Z}\sub{Fe,500}$ for lower-temperature clusters. However, we note that the original F98 data also shows a slight anti-correlation between temperature and $Z\sub{Fe}$, even though they intentionally masked-out the clusters' central regions. In fact, there is a clear anti-correlation present in the original data of most of the cluster samples we consider here, regardless of their choice of observed aperture. And, given that our homogenisation process actually \textit{flattens} the slope of this anti-correlation slightly for all but two of the cluster samples\footnote{The steepness of the $kT_{500}$ - $\bar{Z}\sub{Fe,500}$ anti-correlation for the F98 and T04 increases after homogenisation because these studies adopted a fixed spatial aperture (in both cases, $69-274\,h_{73}^{-1}$kpc), rather than one as a function of each object's scale radius (\eg $r_{500}$). If uncorrected, this leads to an over-estimation of the average metallicity for larger, more extended clusters at low redshift.}, we consider it unlikely that our choice of fixed slope for the iron abundance profile of CC clusters is the dominant cause of this trend.

Alternatively, a temperature-dependent change in the way the iron abundance is derived could be the cause. The presence of an `Fe bias' (\citeauthor{Bu00} 2000a, section A.1) can lead to an over-estimation of $Z\sub{Fe}$ in cooler systems whose spectra are dominated by the Fe L-shell line complex at $\sim 1$ keV, rather than the Fe K-shell line complex at $\sim 6.5$ keV. In the context of clusters, this could lead to higher measured iron abundances for cooler CC systems, if their bright, metal-rich cores have temperatures around 1 keV \citep{N05a}. However, again, the fact that a $T$-$Z\sub{Fe}$ anti-correlation is seen even in data which excludes the central regions (or when considering $M\sub{Fe}/M\sub{gas}$ rather than $Z\sub{Fe}$, \citealt{DG04}) makes it unlikely that an Fe bias is the primary cause here.

Finally, when analysing mock X-ray spectra of four model galaxy clusters, \citet{Ra08} noted that for their cluster in the transition region of $k\bar{T}\sub{ew} \sim 2-3$ keV, neither the Fe-L lines nor the Fe-K lines dominated the spectrum (see also \citealt{Si09}). In such cases, both sets of line complexes contribute to the estimated iron abundance, causing an over-estimation of the true, average ICM iron abundance of up to 20 per cent. Although the peak in the $kT_{500}$ - $\bar{Z}\sub{Fe,500}$ relation for our dataset ($\Temp \sim 0.4$, or $k\bar{T}\sub{ew} \sim 3.7$ keV) is slightly above the transition region identified by \citet{Ra08}, this phenomenon could still be a contributing factor to the anti-correlation we see. However, a reduction in the iron abundance at the peak in the relation by the maximum amount predicted by \citet{Ra08} would still leave a residual negative correlation between $T$ and $Z\sub{Fe}$ for our cluster dataset.

Therefore, we must consider that some of the $T$-$Z\sub{Fe}$ anti-correlation we observe for clusters could be physical. For example, the decrease in iron abundance with temperature for clusters coincides with a similar increase in their baryon fraction (\eg \citealt{Lin03,V06,Gi09,Mc10}). More efficient accretion of pristine gas onto the largest DM haloes would explain both these trends. The peak in the $T$-$Z\sub{Fe}$ relation at $\Temp \sim 0.35$ could therefore indicate a `sweet spot' for clusters, below which feedback processes more efficiently remove metals, and above which infall processes more efficiently dilute the ICM. This is discussed further with regard to our galaxy evolution model in \S \ref{sec:Iron_in_model_clusters}.

\subsection{The $T$-$Z\sub{Fe}$ relation for groups}\label{sec:T-Z for groups}
Below $\Temp = 0.1$, we can see the opposite trend to that seen for clusters -- the mean iron abundance in the hot gas surrounding groups seems to positively correlate with temperature. When assessing if this is a real correlation, we note that the transition between these two trends occurs at a temperature [$\Temp \sim 0.35$] that is 0.25 dex higher than our chosen transition temperature between groups and clusters. This suggests that the different ways we treat groups and clusters in our homogenisation process is not causing the reversal in the $T$-$Z\sub{Fe}$ trend here.

A linear fit to the data in the range $-0.4 < \Temp < 0.1$ yields the following relation for groups:

\begin{equation}\label{eqn:TZR for groups}
	\tn{log}(\bar{Z}\sub{Fe,500}) = (\ObsTZslopeGroups\tn{\scriptsize{$\pm \ObsTZslopeGroupsErr$}})\,\Temp - (\ObsTZnormGroups\tn{\scriptsize{$\pm \ObsTZnormGroupsErr$}})
\end{equation}
with a $1\sigma$ dispersion in $\bar{Z}\sub{Fe,500}$ from residuals of \ObsTZscatterGroups dex.

\citeauthor{Bu00} (2000a, section A.1) has pointed out that the Fe bias mentioned in \S \ref{sec:T-Z for clusters} could cause an under-estimation of the iron abundance for groups relative to clusters. In lower-temperature systems, emission from the cooler Fe L-shell line complex dominates the determination of $Z\sub{Fe}$ from spectral fitting models. Gas at slightly different temperatures will excite Fe-L lines in this complex from $\sim 0.7$ to 1.3 keV, forming a broad emission peak around 1 keV. A simple 1-temperature (1T) model fit to such a spectrum will not be able to reproduce this broad peak, and will inevitably decrease the assumed iron abundance in order to increase the modelled contribution from the flatter bremsstrahlung continuum emission. Such an effect becomes less significant for hotter systems, where the Fe K-shell line complex at $\sim 6.5$ keV dominates instead. This Fe bias could contribute to the decrease in $Z\sub{Fe}$ below $\Temp \sim 0.2$ in our dataset. However, RP09 have shown that both 2T models and mass-weighted measurements also indicate lower iron abundances in groups relative to clusters. It is therefore unlikely that an Fe bias is the sole cause of the $T$-$Z\sub{Fe}$ correlation we see here.

RP09 have argued for a physical explanation, where this correlation is due to more efficient metal ejection from lower-mass systems by SN winds at early times, before the group virialised (see also \citealt{Li16}). Metal-rich gas removal via AGN feedback at later times could also have a contribution (\eg \citealt{K09,Sa16}). The fact that groups are more susceptible to such processes than clusters could also explain the larger scatter in the $T$-$Z\sub{Fe}$ relation for groups seen in Fig. \ref{fig:kT-FeH_Obs}. These physical processes will be discussed further, with reference to our galaxy evolution model, in \S \ref{sec:Iron_in_model_clusters}.

\subsection{An object-for-object comparison}\label{sec:An object-for-object comparison}
There are many systems in our dataset that have been studied by more than one of the observational samples we consider. This allows us to assess whether the homogenised values of $T$ and $Z\sub{Fe}$ we obtain from different studies are in good agreement with each other.

\subsubsection{Comparing temperatures}
In Fig. \ref{fig:Temp_comp_Obs}, we compare the temperatures at $r_{500}$ obtained from different samples for the same objects. All objects with three or more measurements of $T_{500}$ and $\bar{Z}\sub{Fe,500}$ are shown, and are ordered from left to right by ICM temperature. There is very good agreement among the different samples. This close compatibility is partly due to the fact that temperature profiles appear quite similar among groups and clusters (\S \ref{sec:Temperature estimation}). However, it is also an indication that our homogenisation process is working properly.

The high temperatures obtained for three of the clusters from the P03 sample are a clear exception to this close compatibility. For S159-03, Hydra A, and A4059, the derived $T_{500}$ is more than 0.1 dex higher than that obtained from other studies. In each case, this is due to the ambient temperature quoted by P03 being significantly higher than the mean or even peak temperature quoted by other works. Given that we are already considering their ambient temperature values to represent the maximum temperature in the ICM (\S A.1.3), it is difficult to see how these contrasting measurements can be reconciled.

We note here that there is not enough overlap in our dataset between \textit{Chandra} and \textit{XMM-Newton} measurements of the same systems to determine if there is a systematic offset in the estimated temperature or iron abundance from their EPIC and ACIS instruments (see \eg \citealt{An09,Sc15}).

\subsubsection{Comparing iron abundances}
Fig. \ref{fig:FeH_comp_Obs} shows a comparison of the iron abundances obtained from different samples for the same objects. While the compatibility among the different samples is not as tight as for temperature in Fig. \ref{fig:Temp_comp_Obs}, the values of $\bar{Z}\sub{Fe,500}$ obtained are reassuringly similar, and in most cases compatible within the errors. This is promising, given that each sample measured metallicity differently.

It is also encouraging that the iron abundances obtained for groups using our default abundance profile (Eqn. \ref{eqn:Z_profile_DGM01}) are in reasonable agreement with those obtained using individually-measured abundance profiles by RP09 (see also \S A.2.3). This also indicates that our homogenisation method is working well.

Most of the discrepancies seen in Fig. \ref{fig:FeH_comp_Obs} are due to large differences in the iron abundances reported by the original studies. Although our homogenisation process reduces these discrepancies, it cannot completely remove them.

One related and interesting trend is that, in many cases, the predicted value of $\bar{Z}\sub{Fe,500}$ for a given object increases with the recency of the analysis. Two good examples of this are MKW3s and A1795, which have each seen a systematic increase of $> 0.1$ dex in their estimated $\bar{Z}\sub{Fe,500}$ from the year 1998 to 2011. Given that 4 of the 5 samples that contain these clusters used \textit{XMM-Newton} data, and that there is no correlation between the chosen aperture size and the age of the analysis (indeed, F98 and T04 use exactly the same aperture), we presume that this upward revision is due to continuous improvements in the atomic data assumed and the way metallicities are obtained from X-ray spectra.

\section{The galaxy evolution model} \label{sec:The_semi-analytic_model}
In order to study the physical processes causing the trends discussed in \S \ref{sec:Obs_results}, we now turn to the semi-analytic model of galaxy evolution, \textsc{L-Galaxies} \citep{S01,G10,H15}, which is run on the dark matter (DM) subhaloes identified in the \textsc{Millennium} N-body simulation \citep{S05}. Newly-formed DM subhaloes are seeded with hot gas at high redshift in proportion to their virial mass ($M\sub{vir}$). This gas is then allowed to cool, form stars, and be blown back out into the interstellar medium (ISM) and CGM via supernova (SN) feedback. Modelling of gas heating from AGN feedback is also included. \textsc{L-Galaxies} uses analytic prescriptions of physical processes, motivated by observations and simulations, to govern the transfer of mass and energy among seven galactic components (the central black hole, bulge stars, disc stars, halo stars, ISM, CGM/ICM, and ejecta reservoir). The model has been shown to reproduce the Tully-Fisher relation and large-scale clustering of galaxies \citep{G10}, the galaxy stellar mass and optical luminosity functions from $z=0$ to 3 \citep{H13}, and the chemical properties of low-$z$ galaxies \citep{Y13}. We refer the reader to the supplementary material in \citet{H15} for more details on the model.

The \textsc{Millennium} simulation has a particle resolution of $1.18\times 10^{9}\;h_{73}^{-1}\Msun$, and is able to reliably represent the internal structure of DM haloes (otherwise known as friends-of-friends, or FOF, groups) above a resolution of $\sim 1000$ particles, equating to masses above $\sim 1.18\times 10^{12}\;h_{73}^{-1}\Msun$. This means that the density profiles within the group and cluster haloes investigated here, with virial masses above $10^{13} \;h_{73}^{-1}\Msun$, are very well resolved. A WMAP1 cosmology \citep{S03} with the following parameters is assumed in the simulation: $\Omega_{\textnormal{m}}=0.25$, $\Omega_{\textnormal{b}}=0.045$, $\Omega_{\Lambda}=0.75$, $n_{s}=1$, $\sigma_{8}=0.9$, and $\textnormal{H}_{0}=73$ km s$^{-1}\,$Mpc$^{-1}$.

This present work is based on the version of \textsc{L-Galaxies} presented by \citet{Y13}, which is an adaptation of the model discussed in \citet{G10}, including significant improvements to the chemical enrichment modelling. A new galactic chemical evolution (GCE) scheme was implemented, so that the delayed enrichment of eleven individual chemical elements (H, He, C, N, O, Ne, Mg, Si, S, Ca, and Fe) from SNe-Ia, SNe-II, and stellar winds could be properly modelled. This scheme includes the use of mass- and metallicity-dependent stellar yields and lifetimes, and a reformulation of the associated SN feedback so that energy and heavy elements are released into the ISM and CGM when stars die. Such a scheme is an improvement on the `instantaneous recycling approximation', which is sometimes used in galaxy formation models for its simplicity, but does not adequately describe the delayed enrichment of metals from long-lived stars.

The model parameters used in the present work are identical to those used by \citet{Y13}, with the exception of those modifications discussed in \S \ref{sec:Model results}.

\begin{table*}
\centering
\begin{tabular}{ccccccccc}
\hline \hline
$^{1}$Name & $^{2}f\sub{b,cos}$ & $^{3}\kappa\sub{AGN}$ & $^{4}\alpha\sub{IMF}$ & $^{5}A\sub{SNIa}$ & $^{6}\tn{DTD}\sub{SNIa}$ & $^{7}f\sub{hot,SNIa}$ & $^{8}f\sub{hot,SNII}$ & $^{9}f\sub{hot,AGB}$ \\
\hline
Original model & 0.17 & $1.5 \times 10^{-3}$ & 2.3 & 0.028 & power law & 0.0 & 0.0 & 0.0\\
New model & 0.1543 & $7.5 \times 10^{-4}$ & 2.3 & 0.028 & power law & 0.0 & 0.0 & 0.0\\
Extra iron model & 0.1543 &$7.5 \times 10^{-4}$ & 2.15 & 0.04 & power law & 1.0 & 0.8 & 0.0 \\
\hline \hline
\end{tabular}
\caption{The different \textsc{L-Galaxies} model versions considered in this work. \textit{Column 1}: Model name. \textit{Column 2}: Assumed cosmic baryon fraction (either 0.17 from \textit{WMAP1}, or 0.1543 from \textit{Planck}). \textit{Column 3}: Black-hole hot accretion efficiency, in $\Msun$/yr (\citealt{YK14}, eqn. 3). A lower value implies weaker AGN feedback. \textit{Column 4}: Modulus of the high-mass-end slope of the Chabrier IMF used for the GCE. A lower value indicates a more top-heavy IMF. \textit{Column 5}: Fraction of stellar objects in the mass range $3-16\Msun$ assumed to be SN-Ia-producing binary systems. \textit{Column 6}: Form of the SN-Ia delay-time distribution (DTD). For all models considered here, a power law of slope -1.12 is assumed (\citealt{Y13}, section 4.1). \textit{Column 7}: Fraction of ejecta material from SN-Ia in the stellar disc that is assumed to be deposited directly into the hot gas (CGM/ICM). In all cases, stars that die in the stellar bulge are assumed to directly pollute the hot gas. \textit{Column 8}: Same as column 7, but for SN-II ejecta. \textit{Column 9}: Same as column 7, but for AGB wind ejecta.}
\label{tab:Model_setups}
\end{table*}

\section{Model sample} \label{sec:Model sample}
We select a sample of 2456 model DM haloes with $M_{200}\geq 1.0\times10^{13}\,h_{73}^{-1}\Msun$ at $z=0$ from the Munich semi-analytic model of galaxy formation, \textsc{L-Galaxies}. Following our earlier classification of galaxy groups and clusters (\S \ref{sec:Temperature estimation}), 2294 of these DM haloes are defined as hosting a galaxy group, and 162 are defined as hosting a galaxy cluster. Our chosen threshold value of $\Temp=0.1$ distinguishing groups and clusters roughly corresponds to $M_{200}=1.2\times10^{14}\,h_{73}^{-1}\Msun$. Interestingly, this value is close to the mass threshold chosen by \citet{H15} for truncating ram-pressure stripping in group environments. It was found that stripping of hot gas needs to be suppressed below this mass in \textsc{L-Galaxies} in order to match the observed fraction of passive dwarf galaxies at low redshift.

Three variations of the core \textsc{L-Galaxies} model are considered in this study. Our \textit{original model} is that used by \citet{Y13} when studying the chemical composition of the gas and stars within galaxies. Our \textit{new model} contains some improvements to the way infall onto DM haloes is modelled (\S \ref{sec:Infall modifications}). And our \textit{extra iron model} further includes changes to the parameters that define the GCE treatment in the model (\S \ref{sec:Iron_in_model_clusters}). The differences between these three variations are detailed in Table \ref{tab:Model_setups}.

The methods used to calculate $r_{500}$, $T_{500}$, and $\bar{Z}\sub{Fe,500}$ for our model systems are as close as possible to those used for our observational dataset. Our chosen scale radius of $r_{500}$ is calculated as described in \S \ref{sec:Radius estimation}, using the values of $r_{200}$ and $M_{200}$ obtained from the \textsc{Millennium} simulation for each model DM halo. The halo density profiles in the \textsc{Millennium} simulation are known to typically follow an NFW profile \citep{Lud13}. 

$T_{500}$ is obtained by first inverting Eqn. \ref{eqn:temp_profile_V06} (for clusters) or Eqn. \ref{eqn:temp_profile_RP07} (for groups) to determine the mean temperature from the temperature at $r_{200}$. This is given by

\begin{equation}
T_{200} = \mu m\sub{p}\sigma^{2}_{200}/k\;\;, 
\end{equation}
where $\sigma$ is the velocity dispersion, $m\sub{p}$ is the mass of a proton, $\mu m\sub{p}$ is the average mass of the particles (baryons and leptons) in the ICM, $\mu=0.58$, and $k=8.6173\times10^{-8}\,\tn{keV/K}$ is Boltzmann's constant. Then, Eqn. \ref{eqn:temp_profile_V06} or Eqn. \ref{eqn:temp_profile_RP07} are used again to obtain $T_{500}$ from the mean temperature.

The average iron abundance within $r_{500}$ is obtained using the same gas density and iron abundance profiles described in \S \ref{sec:Iron abundance estimation}. We consider all model clusters to have CCs, as they all exhibit AGN feedback at $z=0$ in \textsc{L-Galaxies}. Abundances are also normalised to the solar abundances provided by \citet{GS98}, and are left as mass weighted, in order to fairly compare with the mass-weighted values we derive for our observational dataset (see \citealt{Ra08,Cr13}).

The semi-analytic model does not provide any spatial information on the distribution of baryonic mass within DM haloes. Therefore, for the gas density profiles, we assume that $r\sub{c}=a$ and determine $\beta$ using the fit to the emission-weighted ICM temperature for groups and clusters provided by \citet{Sa03};

\begin{equation}\label{eqn:T-beta}
\beta = 0.439\,\bar{T}\sub{ew}^{\,\,\,0.20}\;\;.
\end{equation}
The trend that gas-density profiles are flatter in lower-temperature systems has been noted by a number of studies (\eg \citealt{ME97,PCN99,HMS99,Sa03}), and is also present in the data compiled from the \citet{RB02} and RP09 samples for our observational dataset. For Eqn. \ref{eqn:T-beta}, the slope reaches the canonical value of 2/3 often assumed for galaxy clusters at $\Temp \sim 0.75$. The consequences of a temperature-dependent gas-density slope on the baryon factions in model groups and clusters are discussed in \S \ref{sec:The baryon fraction}.

\section{Model results} \label{sec:Model results}
\subsection{The baryon fraction} \label{sec:The baryon fraction}
We begin our analysis of our model results by first looking at the baryon fractions ($f\sub{b}$) in groups and clusters. We are careful to define $f\sub{b}$ as it is defined in the observational studies to which we compare. Therefore, we consider $f\sub{b}=M_{\tn{ICM,500}}/M_{500}$, where $M_{\tn{ICM,500}}$ is the total hot gas mass within $r_{500}$, and $M_{500}$ is the total gravitating mass within $r_{500}$.

\begin{figure*}
\centering
\includegraphics[width=0.95\textwidth]{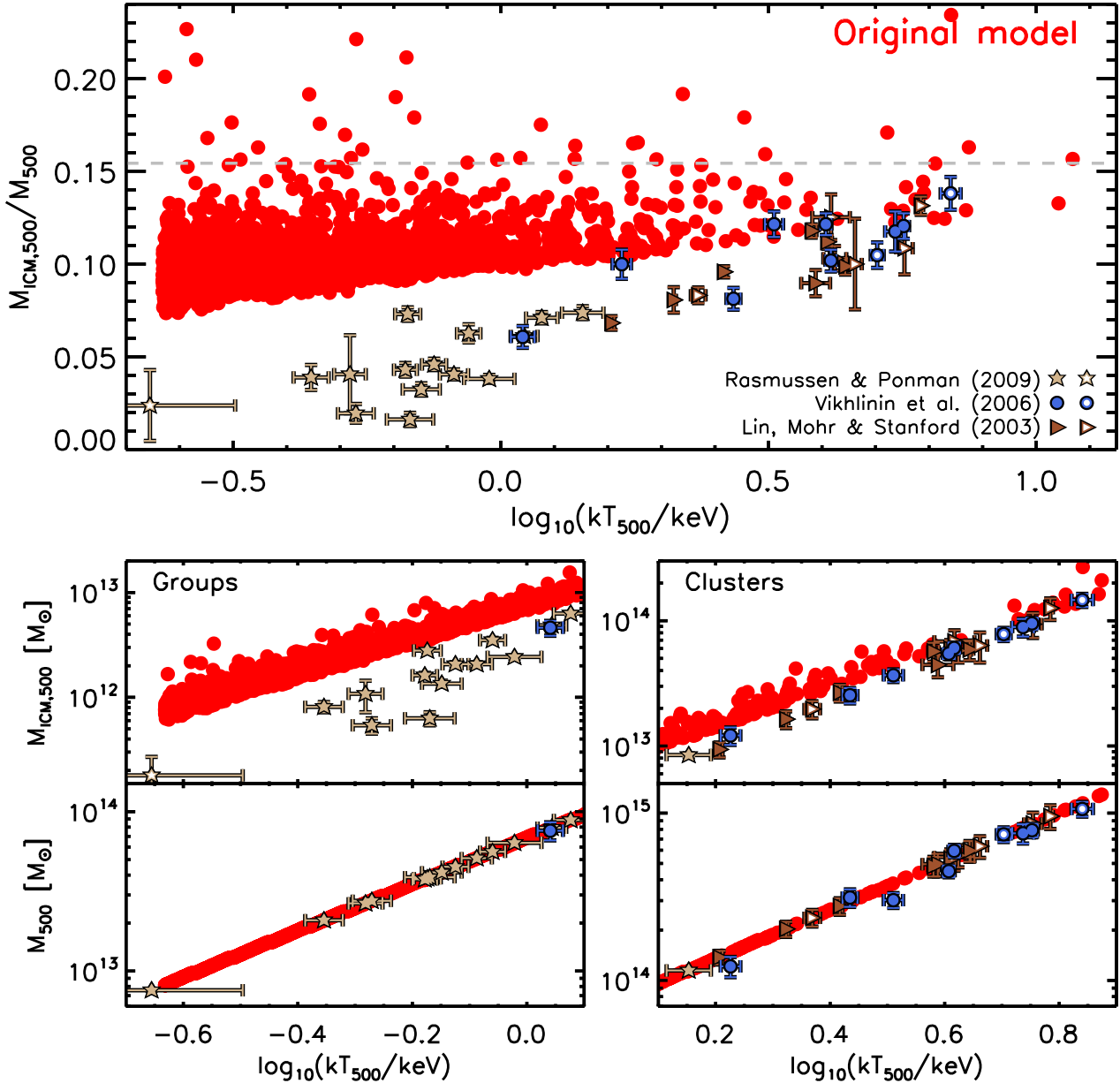} \\
\caption{\textit{Top panel:} The relation between temperature and baryon fraction in the hot ICM of model clusters in our \textit{original model} (red points), before the modifications described in \S \ref{sec:The baryon fraction} are made. This can be compared to Fig. \ref{fig:BaryFrac_After} for our \textit{new model}, after these modifications. The grey, horizontal line indicates the measured cosmic baryon fraction from \textit{Planck}. \textit{Bottom panels:} The ICM gas masses (top row) and DM masses (bottom row) within $r_{500}$ for groups (left column) and clusters (right column). In all panels, observational data from \citet{Lin03}, \citet{V06}, and RP09, also measured within $r_{500}$, are shown for comparison. Filled symbols indicate CC systems and open symbols indicate NCC systems, as defined in \S \ref{sec:Definitions_and_Derivations}.}
\label{fig:BaryFrac_Before}
\end{figure*}

\begin{figure*}
\centering
\includegraphics[width=0.95\textwidth]{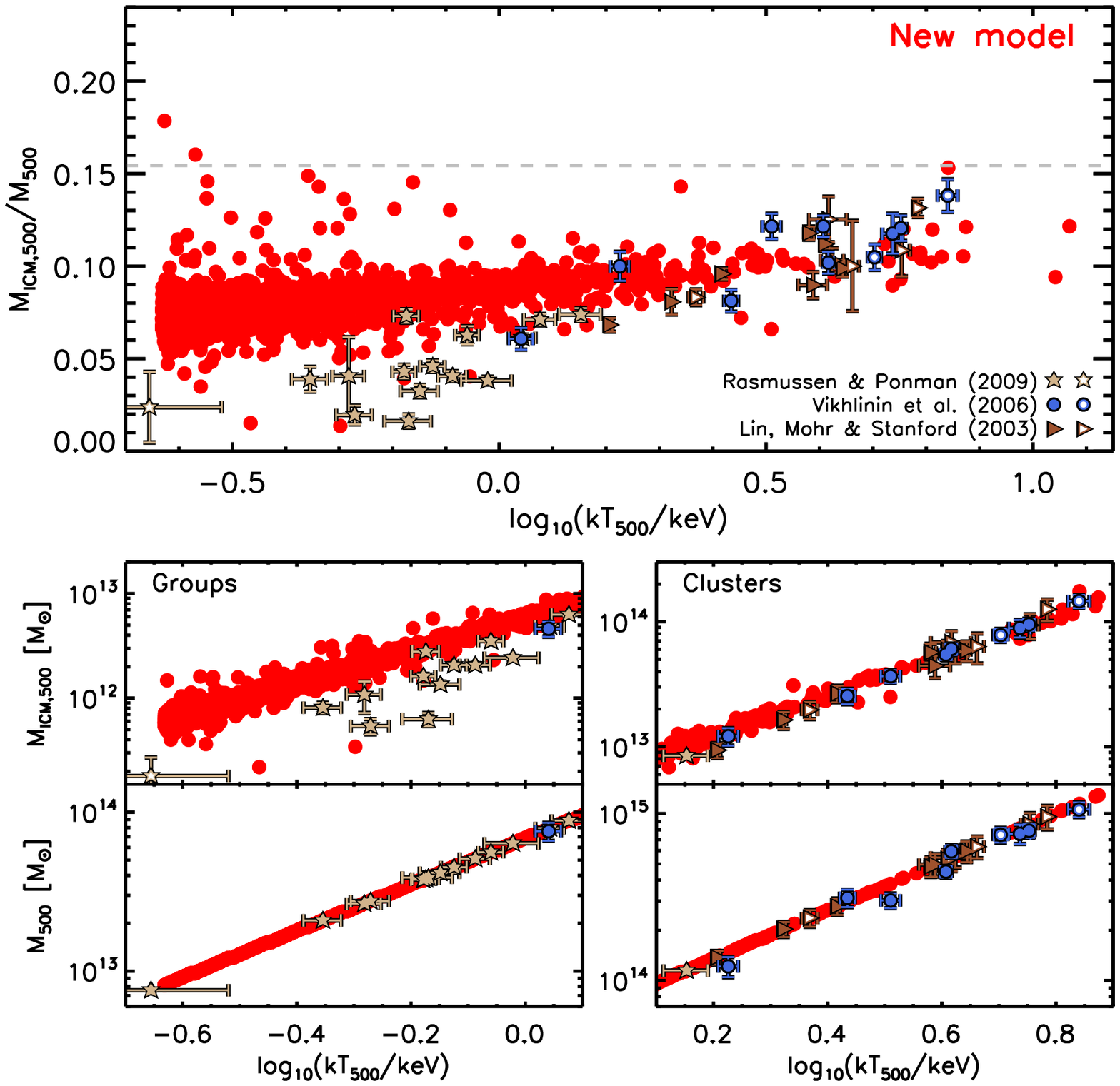} \\
\caption{\textit{Top panel:} The relation between temperature and baryon fraction in the hot ICM of model clusters in our \textit{new model} (red points), after the modifications described in \S \ref{sec:The baryon fraction} are made. This can be compared to Fig. \ref{fig:BaryFrac_Before} for our \textit{original model}, before these modifications. The grey, horizontal line indicates the measured cosmic baryon fraction from \textit{Planck}. \textit{Bottom panels:} The ICM gas masses (top row) and DM masses (bottom row) within $r_{500}$ for groups (left column) and clusters (right column). In all panels, observational data from \citet{Lin03}, \citet{V06}, and RP09, also measured within $r_{500}$, are shown for comparison. Filled symbols indicate CC systems and open symbols indicate NCC systems, as defined in \S \ref{sec:Definitions_and_Derivations}.}
\label{fig:BaryFrac_After}
\end{figure*}

The top panel of Fig. \ref{fig:BaryFrac_Before} shows the $T$-$f\sub{b}$ relation for groups and clusters in our \textit{original model} (red points). That is, the version of \textsc{L-Galaxies} used to study the chemical composition of local galaxies by \citet{Y13}. The same relation for nearby observed systems studied by \citet{Lin03}, \citet{V06}, and RP09 is shown for comparison. The \textit{Planck} value of the cosmic baryon fraction, $f\sub{b,cos}$, is given by the grey dashed line. We note that values of $M_{500}$ for the \citet{Lin03} and RP09 samples have been re-derived here, obtaining $M_{500}$ from a fit to our model $T_{500}$-$M_{500}$ relation, rather than the observed $\bar{T}\sub{ew}$-$M_{500}$ relation of \citet{Fi01b}. This is done so that the differences in the \textit{baryon} content of groups and clusters can be more clearly analysed, without concern for differences in the assumed DM content. The $M_{500}$ values obtained by \citet{V06} are uncorrected, as they are individually calculated for each cluster using a robust hydrostatic equilibrium model of their own. We note that the baryon fractions obtained from the \citet{Lin03} and \citet{V06} data now match each other much more closely. This indicates that the previous discrepancies at fixed temperature were predominantly due to the different assumptions made about the gravitating masses.

Our \textit{original model} $T$-$f\sub{b}$ relation in Fig. \ref{fig:BaryFrac_Before} exhibits a positive correlation (though not as steep as observed). This is because we have allowed the gas density slope to vary with temperature (Eqn. \ref{eqn:T-beta}). Lower-temperature systems are therefore assumed to have flatter gas-density profiles and a smaller fraction of their total hot gas residing within $r_{500}$. Plotting the baryon fraction within $r_{200}$ for our model systems would instead return a relation that is almost independent of temperature, as has been reported in previous theoretical studies (\eg \citealt{N05a,DL04,A10b}). This is the consequence of all DM haloes being `topped-up' to the assumed cosmic baryon fraction (accounting for heating by ultraviolet background radiation) by construction in these models, including \textsc{L-Galaxies}. We note that this is likely to be a poor assumption, as (a) the observed $kT$-$f\sub{b}$ relation extrapolated out to $r_{200}$ is also seen to have a positive gradient (\eg \citealt{Sa03}), and (b) the middle panels in Fig. \ref{fig:BaryFrac_Before} show that there is simply too much gas within $r_{500}$ in our model systems. Indeed, there is a scatter of model groups and clusters with baryon fractions well in excess of $f\sub{b,cos}$ in Fig. \ref{fig:BaryFrac_Before}. Motivated by this fact, we re-assess the way infall is implemented in \textsc{L-Galaxies}.

\subsubsection{Infall modifications}\label{sec:Infall modifications}
We make three modifications to \textsc{L-Galaxies} in order to better model the way pristine gas is allowed to infall onto DM haloes. First, we update the assumed cosmic baryon fraction in the model from the \textit{WMAP1} value of $f\sub{b,cos} = 0.17$ \citep{S03} to the \textit{Planck} value of $f\sub{b,cos} = 0.154$ \citep{PXVI13,Gon13}. This reduces the amount of pristine infall allowed onto central DM haloes, and therefore also $f\sub{b}$. Updating only $f\sub{b,cos}$ from the \textit{WMAP1} to \textit{Planck} value, while assuming all other cosmological parameters are unchanged, is justified in this work because we are interested in more accurately reproducing the baryon fraction in real systems, rather than more accurately representing the masses of their DM haloes.

We also adjust the efficiency of the AGN feedback assumed, noting that its current prescription in \textsc{L-Galaxies} is only to reheat gas, not eject it out of the DM halo. The black hole hot accretion efficiency, $\kappa$, is lowered from $1.5\times10^{-3} \Msun\tn{/yr}$ to $0.75\times10^{-3} \Msun\tn{/yr}$. This ensures that the decrease in $f\sub{b,cos}$ doesn't lead to an under-production of massive galaxies by $z=0$ in our model. We consider this small reduction reasonable, as its previous value was itself tuned to the high-mass end of the $z=0$ stellar mass function (\citealt{G10}, section 3.9). 

The second modification we make is related to the virial mass (\ie $M_{200}$) of the DM halo (\ie FOF group) containing the central cluster galaxy. We have found that $M_{200}$ can gradually vary over time for some model systems, due to changes in the morphology of the cluster. A galaxy cluster's FOF group can be stretched and distorted during interactions with other FOF groups, which in turn affects the value of $r_{200}$. This prompts the infall of pristine gas onto the halo, in order to maintain the baryon fraction at around the value of $f\sub{b,cos}$. Subsequently, the FOF group of these model clusters starts to shrink again, decreasing $M_{200}$ and therefore increasing the baryon fraction above the cosmic limit. We note that the model clusters' total DM-particle mass including DM particles \textit{outside} of $r_{200}$ is not affected in the same way as their $M_{200}$ value, demonstrating that the change in $M_{200}$ is due to changes in morphology, rather than significant accretion or loss of dark matter. Such events are causing the baryon fraction to be over-estimated in some model clusters due to superfluous infall of pristine gas from the IGM.

We have addressed this issue by requiring that a cluster's virial mass cannot decrease with cosmic time. This is done as a pre-processing step before the semi-analytic model is run, by stepping through the DM halo merger trees from low to high redshift, correcting $M_{200}$ where necessary.

\begin{figure}
\centering
\begin{tabular}{c}
\includegraphics[width=0.45\textwidth]{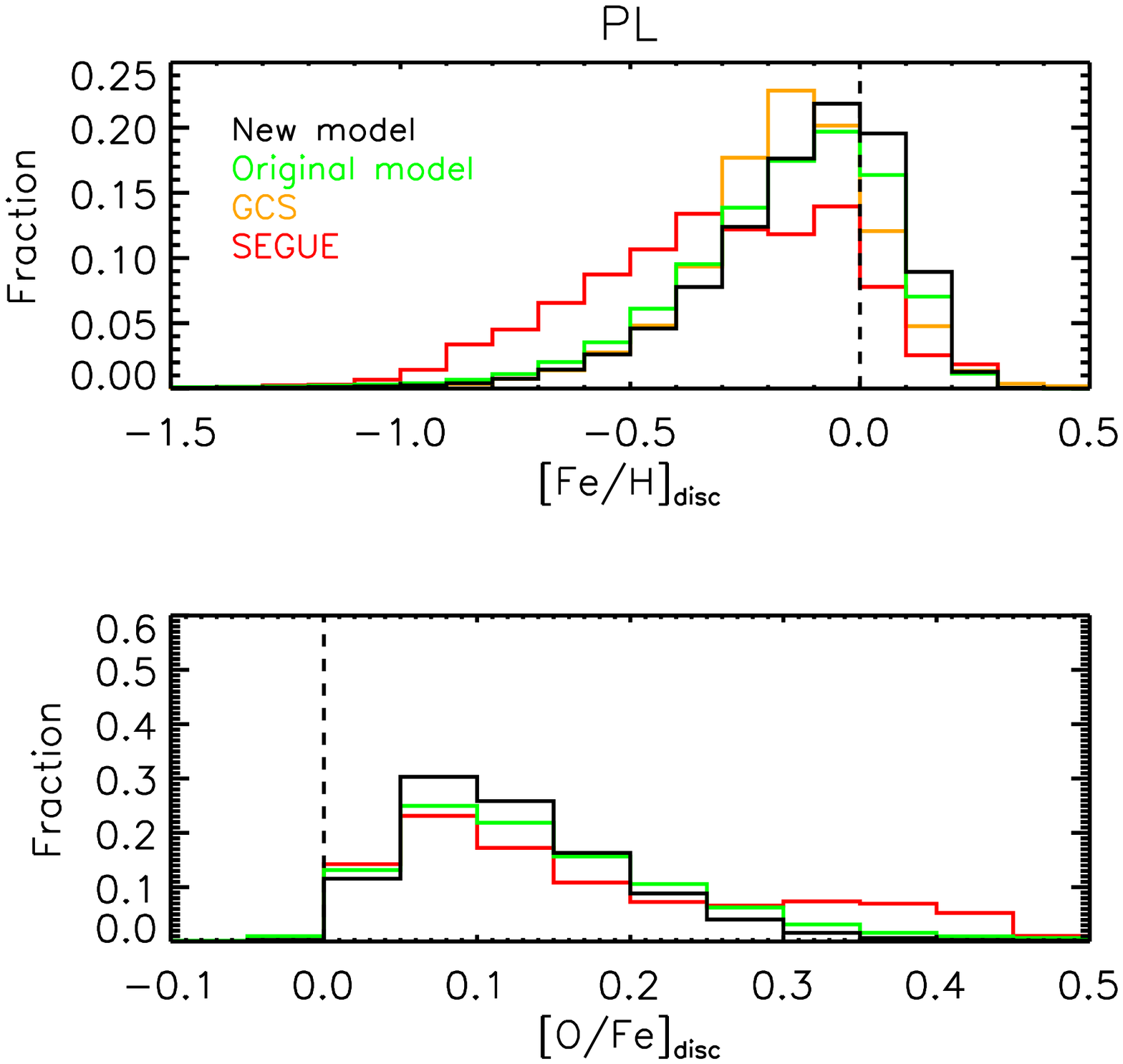} \\
\includegraphics[width=0.45\textwidth]{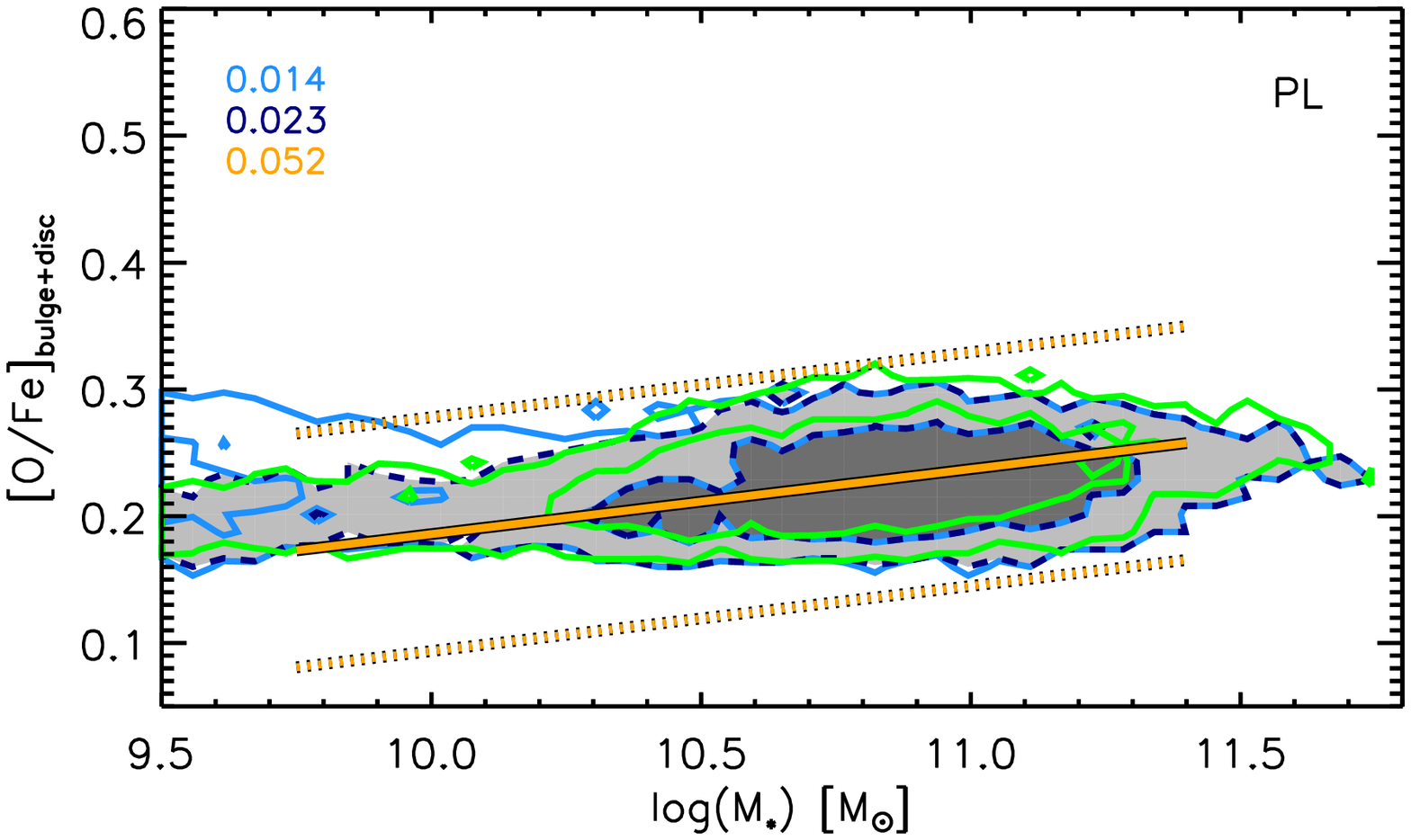}\\
\end{tabular}
\caption{\textit{Top panel:} The [Fe/H] distribution for the stellar discs of Milky-Way-type galaxies at $z=0$ in our \textit{original model} (green) and \textit{new model} (black). \textit{Middle panel:} The [O/Fe] distribution for the same model Milky-Way-type galaxies. Observational data from \citet{HNA09} (yellow) and \citet{B12a,B12b} (red) are shown for comparison (\citealt{Y13}, section 6.2.1). \textit{Bottom panel:} The $M_{*}$-[O/Fe] relation for the stellar components of elliptical galaxies at $z=0$ in our \textit{original model} (green) and \textit{new model} (filled contours). Observational data from \citet{JTM12} are shown for comparison (\citealt{Y13}, section 6.3.2).}
\label{fig:MW_FeH_OFe_NewModel}
\end{figure}

The third modification we make also affects the amount of baryonic infall allowed onto DM haloes. Previously, when calculating the amount of infall required, only baryons in satellite galaxies \textit{within} $r_{200}$ were considered, and the baryon fraction was assumed to be $M\sub{b,200}/M_{200}$ when calculating infall. Such a choice, although reasonable, does not take account of the large number of FOF-group-member satellite galaxies that will fall within $r_{200}$ of the central object at some later time. As a satellite approaches $r_{200}$ in our model, its DM subhalo is already being stripped, but \textsc{L-Galaxies} does not allow its baryons to be stripped until it falls into the cluster. Therefore, when a large satellite does cross $r_{200}$, it does so with an enhanced baryon fraction. This causes a jump in the baryon fraction of the cluster above $f\sub{b,cos}$, as it is has already been `topped-up' to the cosmic baryon fraction by pristine infall.

To remedy this issue, we instead check that the value of $M\sub{b,FOF}/M_{200}$, rather than $M\sub{b,200}/M_{200}$, does not exceed $f\sub{b,cos}$ when calculating infall. This larger value therefore accounts for \textit{all} baryons that are in the FOF group, including those not currently within $r_{200}$, and prevents erroneous accretion of pristine gas before the infall of a new DM subhalo.

Note that both $M\sub{b,200}/M_{200}$ and $M\sub{b,FOF}/M_{200}$ are alternative interpretations of the baryon fraction to $M\sub{ICM,500}/M_{500}$, which is what is typically measured when observing of the hot, X-ray-emitting gas in groups and clusters. For this reason, we always plot $M\sub{ICM,500}/M_{500}$ when comparing to observations in this work.

Fig. \ref{fig:BaryFrac_After} shows the effect of the three modifications described above on the baryon fractions of our model groups and clusters. Nearly all systems now have $f\sub{b}$ below $f\sub{b,cos}$ in this \textit{new model}, and the baryon fractions in clusters are now in better agreement with those observed. This is due to our \textit{new model} matching the observed ICM masses, even in low-temperature clusters (middle-right panel, Fig. \ref{fig:BaryFrac_After}).

However, the ICM masses in galaxy groups are still over-estimated in \textsc{L-Galaxies} (middle-left panel, Fig. \ref{fig:BaryFrac_After}). This is a strong indication that gas removal by feedback is also required in our model. As mentioned in \S \ref{sec:T-Z for groups}, AGN feedback is a good candidate for this (\eg \citealt{Bo08,Fa10}), as it is likely to have a more significant effect in the shallower gravitational potential wells of groups than those of clusters, while also not affecting even smaller systems which don't contain super-massive black holes (SMBHs). Possible improvements to our AGN feedback modelling are discussed further in \S \ref{sec:Iron_in_model_groups}.

\begin{figure*}
\centering
\includegraphics[width=0.95\textwidth]{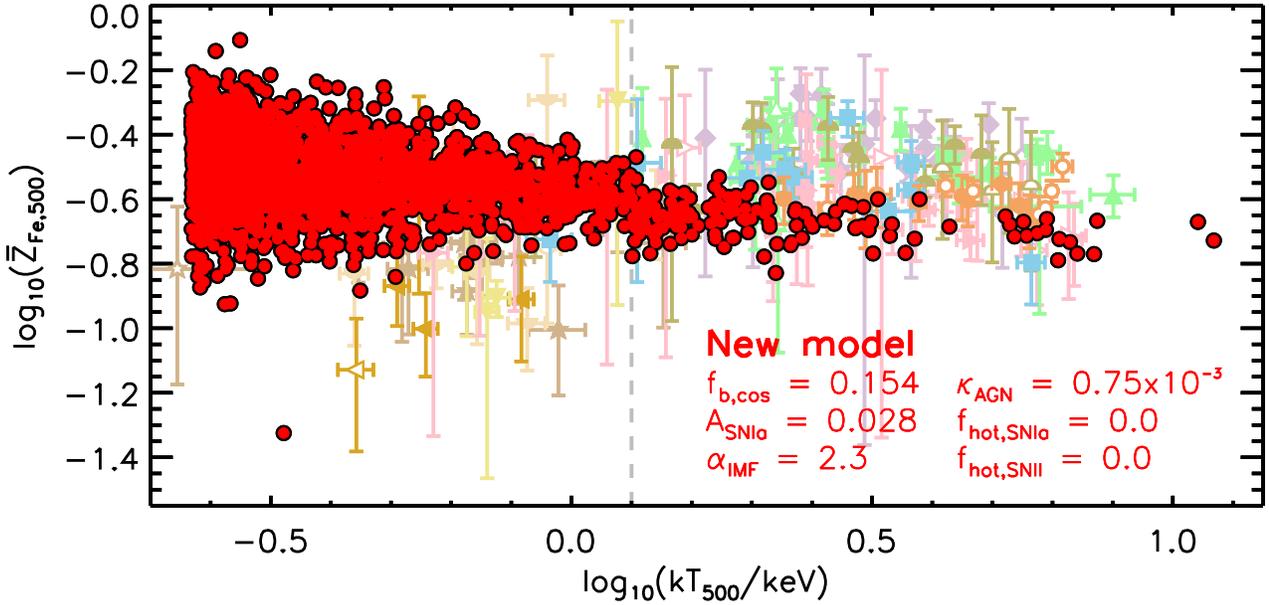}
\caption{The $kT_{500}$-$\bar{Z}\sub{Fe,500}$ relation for model clusters (red points), after the modifications described in \S \ref{sec:The baryon fraction} are implemented. Our full observational dataset is plotted in the background for comparison. The key model parameters for our \textit{new model} are quoted in red (see also Table \ref{tab:Model_setups}).}
\label{fig:kT-FeH_NewModel}
\end{figure*}

\begin{figure*}
\centering
\includegraphics[width=0.95\textwidth]{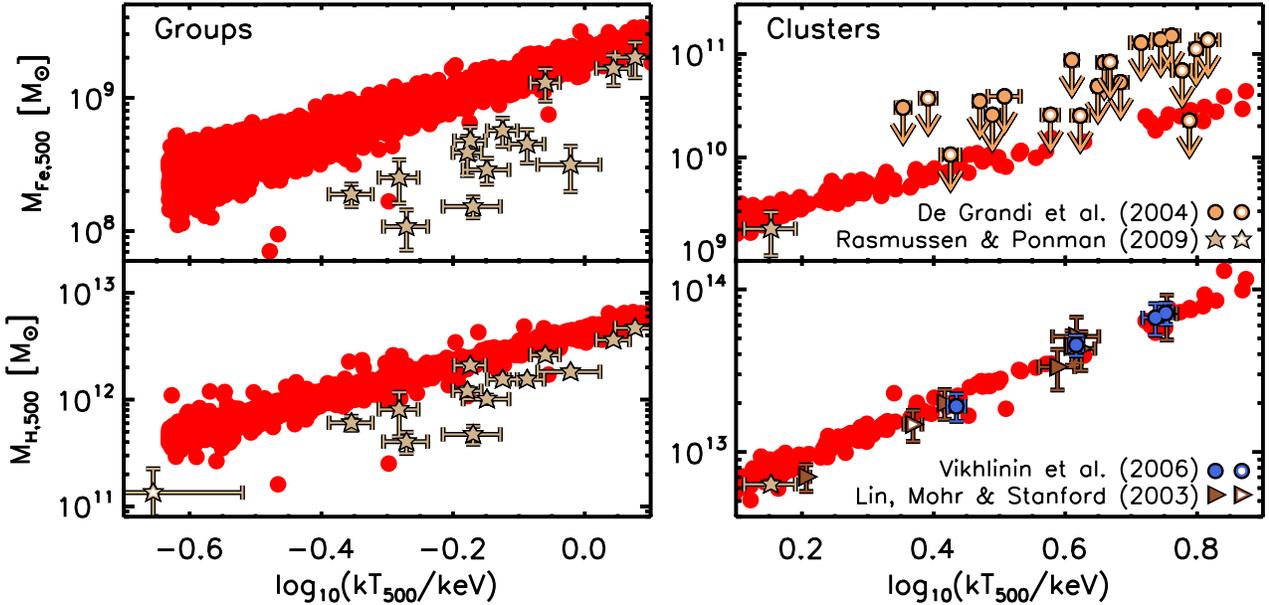}
\caption{The ICM iron masses (top row) and hydrogen masses (bottom row) within $r_{500}$ for groups (left column) and clusters (right column) for our model sample (red). In all panels, observational data, also measured within $r_{500}$, are shown for comparison. Filled symbols indicate CC systems and open symbols indicate NCC systems, as defined in \S \ref{sec:Definitions_and_Derivations}. Hydrogen mass estimates for the three observational samples have been obtained from the total gas masses provided, assuming a solar abundance of helium and ICM metallicites measured for each system: $M\sub{H} = M\sub{ICM}(1-Y\asun-(Z\asun\cdot{}\bar{Z}_{500}))$. For the RP09 sample, the total metallicities measured by \citet{RP07} were used. For the \citet{Lin03} and \citet{V06} samples, the mean of the iron abundances obtained for the same clusters from our observational dataset were used.}
\label{fig:kT-MFe_MH_NewModel}
\end{figure*}

\subsection{Metals in galaxies}\label{sec:Metals in galaxies}
Before turning to the iron abundances in our model groups and clusters, it is first important to check that the modifications described in \S \ref{sec:The baryon fraction} do not destroy \textsc{L-Galaxies'} correspondence with observations on smaller scales. Therefore, we show the [Fe/H] and [O/Fe] distributions for the stellar discs of model Milky-Way-type galaxies and the $M_{*}$-[O/Fe] relation for the stars in local elliptical galaxies in Fig. \ref{fig:MW_FeH_OFe_NewModel}.

We can see that the peaks of the new Milky Way [Fe/H] and [O/Fe] distributions (black) are shifted very slightly towards solar values (vertical, dashed lines in Fig. \ref{fig:MW_FeH_OFe_NewModel}) compared to our \textit{original model} (green), and the oxygen enhancements in the most massive ellipticals are shifted to slightly lower values (by $\sim 0.01$ dex). These small changes are likely due to a slight reduction in galaxy star-formation rates at high redshift that is a consequence of a lower assumed cosmic baryon fraction. However, overall, our modifications have had a negligible effect on the chemical properties of the stars within galaxies. The effect of our \textit{extra iron model} on the chemical composition within galaxies is discussed in \S \ref{sec:Iron_in_model_clusters}.

\subsection{Iron in model clusters}\label{sec:Iron_in_model_clusters}
In Fig. \ref{fig:kT-FeH_NewModel}, we show the $kT_{500}$-$\bar{Z}\sub{Fe,500}$ relation for all our 2456 model systems at $z=0$ (red), with our full observational dataset plotted in the background for reference. $\bar{Z}\sub{Fe,500}$ is calculated for our model clusters by rescaling the ratio of the total iron mass to hydrogen mass in the hot gas components of all cluster members within $r_{200}$, using the same process utilised for our observational dataset (\S \ref{sec:Iron abundance estimation}).

Firstly, we note that our \textit{new model} roughly reproduces the iron abundances measured for the hottest ($\Temp \gtrsim 0.8$) and coldest ($\Temp \lesssim 0.2$) clusters. This is partly due to the 0.1 dex increase in iron abundance compared to our \textit{original model}, roughly half of which is due to adopting the \textit{Planck} cosmic baryon fraction, and half due to our other infall modifications (\S \ref{sec:The baryon fraction}). Scaling $Z\sub{Fe}$ to the same radius for the model and observations also has a significant effect, revealing the model to be a better representation of the data than previously thought.

However, intermediate-temperature clusters ($0.2 < \Temp < 0.8$) in our model still appear under-abundant in iron compared to the data by up to $\sim 0.2$ dex. This is because, although the slope of the model $T$-$Z\sub{Fe}$ relation is also negative, it is not as steep as observed (although it does steepen at higher redshift, see \S \ref{sec:Evo}), with a value of \ModelTZslope at $z=0$ compared to \ObsTZslope for our observational dataset.

Agreement between the model and data would improve if the negative slope in the observed $T$-$Z\sub{Fe}$ relation were partly a bias effect (\S \ref{sec:T-Z for clusters}). The true iron abundances of intermediate-temperature systems would then be lower, in better correspondence with \textsc{L-Galaxies}, and the slopes of the model and observed $T$-$Z\sub{Fe}$ relations would be more similar. We also note that the iron \textit{masses} in our model clusters are marginally consistent with the upper limits derived by \citet{DG04} (top-right panel of Fig. \ref{fig:kT-MFe_MH_NewModel}). However, it is also possible that there is simply not enough iron produced and distributed into the ICM in our model systems. 

The problem of low iron abundances in the ICM has been encountered by galaxy evolution models before. For example, \citet{N05a} attempted to boost the iron mass in the ICM of clusters in the \textsc{Galform} semi-analytic model \citep{C00} by assuming a flat stellar IMF for stars formed in starbursts. This allows a much larger fraction of iron-producing SNe-Ia in the stellar populations of central galaxies, which undergo many mergers as they evolve. Alternatively, \citet{A10b} allowed 80 per cent of the metal-rich material ejected by stellar winds and SNe to be deposited directly into the CGM around their model galaxies. The number of SN-Ia progenitor systems was also increased, by altering the IMF and increasing the number of SN-Ia progenitors per stellar population. These changes boost the production of iron as well as its ability to enrich the hot, diffuse gas surrounding galaxies.

\begin{figure}
\centering
\includegraphics[width=0.45\textwidth]{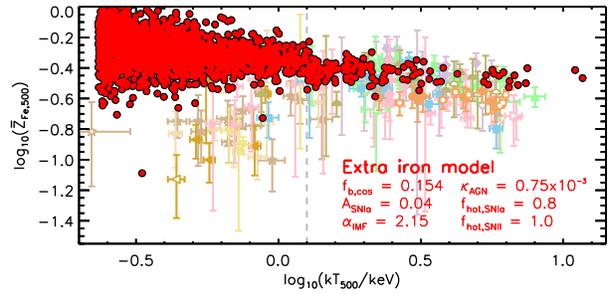}
\caption{The $kT_{500}$-$\bar{Z}\sub{Fe,500}$ relation for model clusters (red) in our \textit{extra iron model}, \ie a set-up designed to enhance the enrichment of the ICM by galactic SNe-Ia. There is better agreement between the model and observations for intermediate-temperature clusters, compared to our \textit{new model}. However, there is worse agreement elsewhere (\S \ref{sec:Iron_in_model_clusters}).}
\label{fig:kT-FeH_Arrigoni_setup}
\end{figure}

\begin{figure}
\centering
\begin{tabular}{c}
\includegraphics[width=0.45\textwidth]{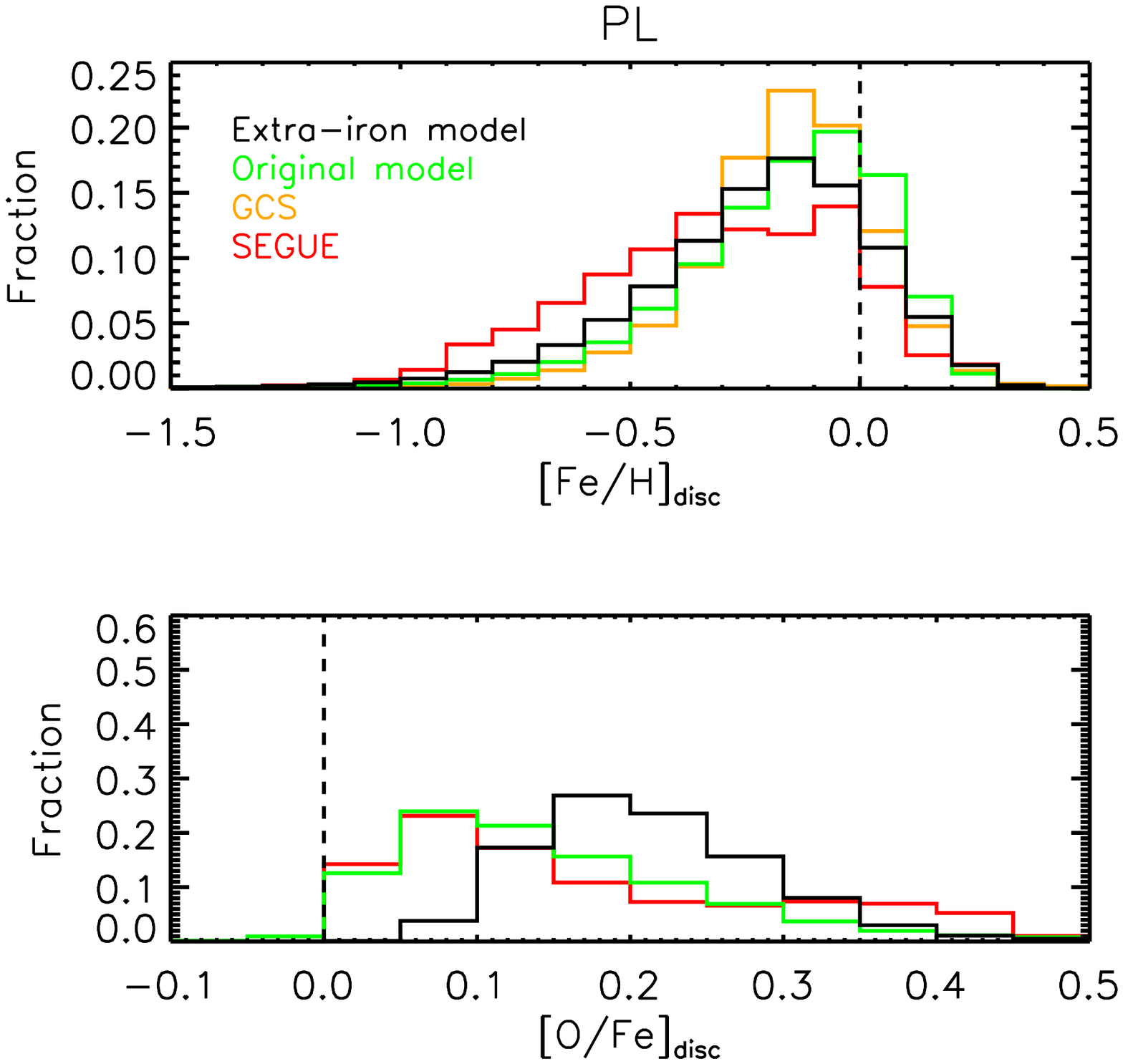} \\
\includegraphics[width=0.45\textwidth]{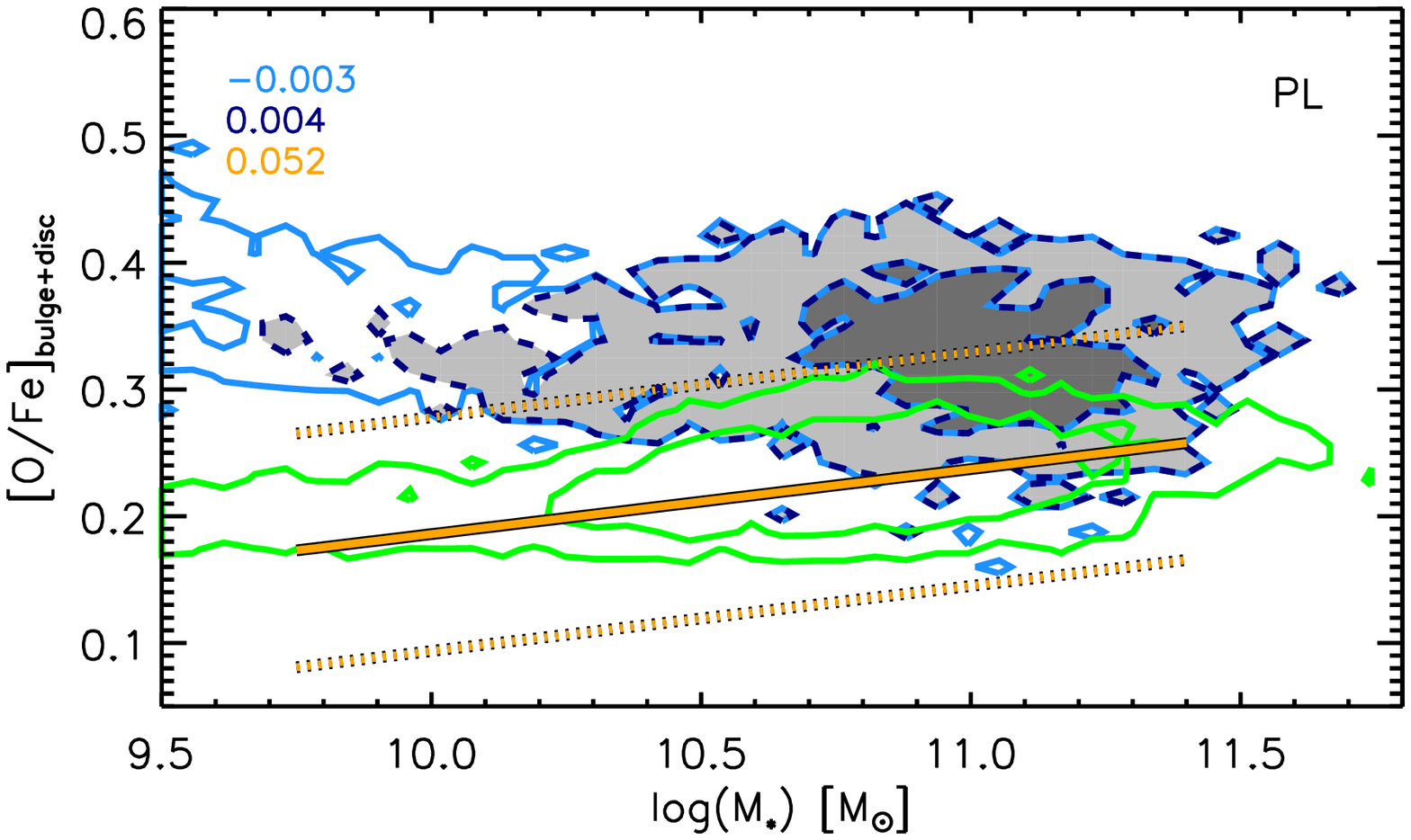}\\
\end{tabular}
\caption{\textit{Top panel:} The [Fe/H] distribution for the stellar discs of Milky-Way-type galaxies at $z=0$ in our \textit{original model} (green) and \textit{extra iron model} (black). \textit{Middle panel:} The [O/Fe] distribution for the same model Milky-Way-type galaxies. Observational data from \citet{HNA09} (yellow) and \citet{B12a,B12b} (red) are shown for comparison. \textit{Bottom panel:} The $M_{*}$-[O/Fe] relation for the stellar components of elliptical galaxies at $z=0$ in our \textit{original model} (green) and \textit{extra iron model} (filled contours). Observational data from \citet{JTM12} are shown for comparison.}
\label{fig:MW_FeH_OFe_Arrigoni_setup}
\end{figure}

Fig. \ref{fig:kT-FeH_Arrigoni_setup} shows the effect of making changes to \textsc{L-Galaxies} similar to those chosen by \citet{A10b}. We set the IMF slope to $\alpha\sub{IMF}=2.15$ (shallower than the original 2.3) and increase SN-Ia production by setting $A\sub{SNIa} = 0.04$ (greater than our default choice of 0.028). This means that 4.0 per cent of stars in the mass range $3-16\Msun$ are assumed to be born as SN-Ia-producing binaries. Given the slightly top-heavier IMF, this equates to 0.16 per cent of \textit{all} stellar objects being SN-Ia progenitors, compared to the 0.11 per cent assumed in our \textit{new model}. We also allow 100 per cent of SN-Ia ejecta and 80 per cent of SN-II ejecta to directly pollute the CGM around galaxies. Hereafter, we refer to this version of the model (which also includes the modifications described in \S \ref{sec:The baryon fraction}) as the \textit{extra-iron model} (see Table \ref{tab:Model_setups}).

We can see from Fig. \ref{fig:kT-FeH_Arrigoni_setup} that the \textit{extra iron model} produces iron abundances more in line with those observed in intermediate-temperature clusters. Although this could be deemed a success (assuming the measured $Z\sub{Fe}$ for these objects is accurate), it is important to note that such changes to the stellar IMF, SN-Ia production rate, and metallicity of galactic winds will also affect the chemical composition of systems smaller than clusters. For example, Fig. \ref{fig:kT-FeH_Arrigoni_setup} also shows that $Z\sub{Fe}$ in groups is now much higher than observed, even more so than in our \textit{new model}. Furthermore, Fig. \ref{fig:MW_FeH_OFe_Arrigoni_setup} shows that the [O/Fe] distribution in Milky-Way-type stellar discs (black, middle panel) is now shifted significantly to higher values compared to our \textit{original model} (green), with a peak around $\tn{[O/Fe]}=0.18$ in contrast to the observed peak at $\sim 0.08$ dex (red). Likewise, the bottom panel of Fig. \ref{fig:MW_FeH_OFe_Arrigoni_setup} shows there is no longer any correspondence with the observed $M_{*}$-[O/Fe] relation for elliptical galaxies (orange) in the \textit{extra iron model}. The amplitude is too high, and there is now a negative correlation between $M_{*}$ and [O/Fe]. \citet{N05b} also found such a negative correlation when implementing their flat starburst IMF.

These inconsistencies highlight the fact that we must be careful when changing GCE parameters in galaxy evolution models. Although altering the stellar IMF, SN-Ia production efficiency, or galactic wind metallicity are promising ways to improve the chemical properties in clusters, they can easily destroy the correspondence between model and data for other systems. Given this, we choose to focus on our \textit{new model} in the rest of this work, which is able to reproduce the iron abundances in the hottest clusters simultaneously with the chemical compositions of (a) the star-forming gas in local emission-line galaxies, (b) the Milky Way stellar disc, and (c) the integrated stellar populations of nearby ellipticals \citep{Y13}.

\subsection{Iron in model groups}\label{sec:Iron_in_model_groups}
In Fig. \ref{fig:kT-FeH_NewModel}, we have shown that the hot gas surrounding galaxy groups is too iron-rich in our \textit{new model}. Observations suggest a strong positive correlation between temperature and $Z\sub{Fe}$ in groups, whereas the slight anti-correlation seen in clusters simply continues to lower temperatures in \textsc{L-Galaxies}.

The left-hand panels of Fig. \ref{fig:kT-MFe_MH_NewModel} tell us that this over-abundance in the hot gas of model groups is due to an excess of iron. This, combined with the fact that these systems also have a slight excess of hydrogen (see bottom-left panels in Figs. \ref{fig:BaryFrac_After} and \ref{fig:kT-MFe_MH_NewModel}) strongly suggests that iron-rich material needs to be removed from their DM haloes. This would simultaneously correct both their baryon fractions and their iron abundances. Metals are already driven out of group-sized DM haloes at high redshift by SN feedback in our model, but this seems to be insufficient. As mentioned in \S \ref{sec:The baryon fraction}, AGN feedback is a promising alternative candidate.

Currently, \textsc{L-Galaxies} does not include a prescription for gas removal via AGN feedback. Instead, only heating of the gas in the ISM and CGM is considered, in order to offset cooling in massive DM haloes. \citet{Bo08} have proposed an AGN-feedback implementation for the \textsc{Galform} semi-analytic model in which AGN can heat \textit{and remove} X-ray-emitting (\ie cooling) gas from the ICM. Given that this gas tends to be at lower radii and more metal rich, such an implementation is likely to also have the desired affect on ICM iron abundances in \textsc{L-Galaxies}. We would also expect to see a peak in the $T$-$Z\sub{Fe}$ relation at intermediate temperatures with such an implementation, as indicated by observations (\S \ref{sec:T-Z for clusters}). This peak would signify the optimum size of DM halo in which the ICM can be most efficiently enriched. An investigation into alternative methods of implementing AGN feedback will be the focus of future work.

\subsection{Iron abundance evolution with redshift}\label{sec:Evo} 
\begin{figure}
\centering
\includegraphics[width=0.45\textwidth]{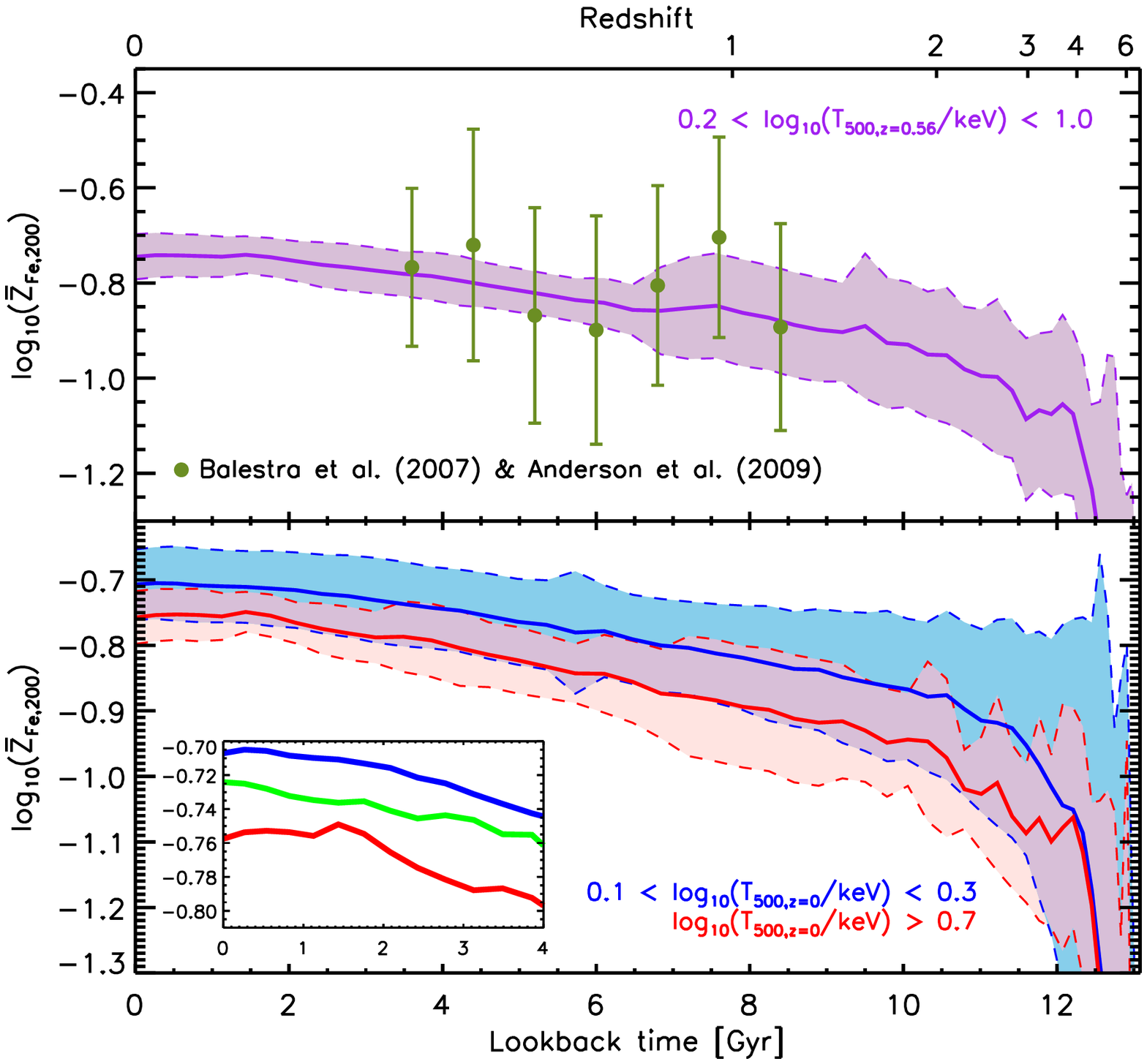}
\caption{\textit{Top panel:} The mean $\bar{Z}\sub{Fe,200}$ evolution for model clusters with $0.2 < \Temp < 1.0$ at $z=0.56$ (23 systems, purple). The $1\sigma$ spread in the mean is given by the light purple region. Observed values of $\bar{Z}\sub{Fe,200}$ for 74 clusters (a total of 80 measurements) with the same temperature range and median redshift from \citet{Ba07} and \citet{An09} are shown for comparison (green points). These are binned in bins of 0.8 Gyr width, with the $1\sigma$ spread in each bin given by the green error bars. \textit{Bottom panel:} The mean $\bar{Z}\sub{Fe,200}$ evolution for the coldest model clusters [$0.1 < \tn{log}(T_{500,z=0}/\tn{keV}) < 0.4$, 125 systems, blue] and the hottest model clusters [$\tn{log}(T_{500,z=0}/\tn{keV}) > 0.7$, 15 systems, red]. Shaded areas indicate the $1\sigma$ scatter in the mean. The inlaid panel shows the same evolution up to lookback time $=4$ Gyr, including intermediate-temperature clusters [$0.4 < \tn{log}(T_{500,z=0}/\tn{keV}) < 0.7$, 21 systems, green].}
\label{fig:FeH_evos_NewModel}
\end{figure}

In the top panel of Fig. \ref{fig:FeH_evos_NewModel}, we show the \textit{mean} evolution of the iron abundance within $r_{200}$ for model clusters from $z=7$ (lookback time $\sim 13\,\tn{Gyr}$) to the present day. 80 measurements of the ICM iron abundance in observed clusters over the redshift range $0.3 < z < 1.27$ from \citet{Ba07} and \citet{An09} are also shown, binned by lookback time (green points). In Fig. \ref{fig:FeH_evos_NewModel}, we only consider model clusters within the temperature range $0.2 < \tn{log}(T_{500}/\tn{keV}) < 1.0$ at $z=0.56$. This is the range of temperatures covered by the observational dataset, which has a median redshift of 0.56. Observed abundances have been corrected in the same way as described in \S \ref{sec:Iron abundance estimation}, assuming the locally-measured iron abundance gradient for CC clusters by \citealt{M11}, although we note that abundance gradients could evolve over time (\eg \citealt{Co06,M16}). Estimates of $r_{200}$ were obtained following \citet{Ev96}, via

\begin{align}
\nonumber r_{200} \;= & \;2.53\,\sqrt{\bar{T}\sub{ew}/10.0\,\tn{keV}}\\
 & \cdot{}(\Omega_{0}[1+z]^{3}+1-\Omega_{0})^{-1/2}\;h_{73}^{-1}\tn{Mpc}\;\;,
\end{align}
where $\Omega_{0} \equiv \Omega\sub{m,z=0} = 0.25$ for our chosen cosmology.

There is clear consistency between our model and the observations across the full comparable redshift range, although the mean scatter in the observed iron abundance measurements is considerable ($\pm0.28$ dex). This agreement is rather surprising, given that the ISM in model star-forming galaxies appears to be over-enriched with oxygen at high redshift, already reaching present-day values by $z\sim2$ (\citealt{YKG12}, section 6.4). Nonetheless, it is promising that \textsc{L-Galaxies} is consistent with the observed evolution of iron in the ICM of hot clusters back to $z=1.27$, given the uncertainties.

Our model indicates that an average of 3 per cent of the \textit{iron mass} found in the ICM of hot clusters at $z=0$ is already present by $z=2$, 17 per cent is present by $z=1$, and 46 per cent is present by $z=0.5$. Given that the accretion of hydrogen onto DM haloes is also occurring at the same time as iron enrichment, this equates to an \textit{iron abundance} fraction ($Z\sub{Fe,z}/Z\sub{Fe,z=0}$) of 66 per cent at $z=2$, 79 per cent at $z=1$, and 85 per cent at $z=0.5$. Our model therefore supports the conclusion that a significant amount of the metallicity evolution in clusters is complete by $z\sim 1$, as suggested by a number of observational studies (\eg \citealt{ML97,AF98,To03,Ba12,M16}).

Our model also indicates that a negative correlation between temperature and iron abundance for clusters has been in place since $z\sim 3$. The bottom panel of Fig. \ref{fig:FeH_evos_NewModel} shows the mean evolution of $\bar{Z}\sub{Fe,200}$ for a set of the hottest (red) and coldest (blue) clusters in our model. The inlaid panel shows the same evolution below a lookback time of 4 Gyr, including intermediate-temperature clusters (green). Colder clusters clearly have a higher iron abundance than hotter clusters. In \S \ref{sec:T-Z for clusters}, we have discussed observational evidence for such a $T$-$Z\sub{Fe}$ anti-correlation at low redshift. At higher redshift, \citet{To03} and \citet{Ba07} have also found a significant negative slope from $z=0.3$ to 1.27 (but see \citealt{Ba12}).

The cause of this weak negative correlation in \textsc{L-Galaxies} is the presence of large hydrogen reservoirs in hotter clusters. At $z=0$, the coldest clusters have, on average, 14 per cent of the iron found in the hottest clusters, but only 12.5 per cent of the hydrogen. This leads to a difference in the mean present-day ICM iron abundance of $\sim 0.05$ dex. At $z\sim 3$, the difference is even greater, around 0.12 dex.

\begin{figure}
\centering
\includegraphics[width=0.45\textwidth]{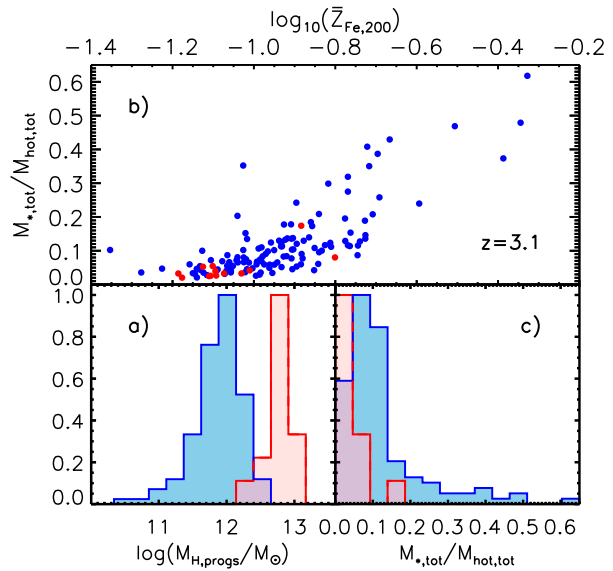}
\caption{\textit{Panel A}: The total CGM hydrogen mass distribution at $z=3.1$ for all minor progenitor systems (\S \ref{sec:Evo}) of the hottest clusters (red) and coldest clusters (blue) in \textsc{L-Galaxies}. All the hot gas present in these minor progenitors will end-up in the cluster's ICM by $z=0$. \textit{Panel B}: The $Z\sub{Fe}$-$M_{*,\tn{tot}}/M\sub{hot,tot}$ relation for model clusters at $z=3.1$. All the stars and hot gas present in all clusters members within $r_{200}$ at this redshift are considered. \textit{Panel C}: The $M_{*,\tn{tot}}/M\sub{hot,tot}$ distribution at $z=3.1$ for the same hottest clusters (red) and coldest clusters (blue).}
\label{fig:SatHists_NewModel}
\end{figure}

Hotter clusters have enhanced hydrogen masses at high redshift because they host a larger number of satellite systems. The hydrogen in the CGM of these satellites is efficiently stripped into the ICM over time through ram-pressure and tidal effects. Panel A in Fig. \ref{fig:SatHists_NewModel} shows that the minor progenitor systems\footnote{The term `minor progenitors' refers to all DM subhaloes that will have merged with the cluster's central DM subhalo by $z=0$. The central DM subhalo itself at any given redshift is referred to as the `main progenitor'.} of the hottest clusters at $z=3.1$ (red) contain a larger mass of hydrogen in their combined CGM than the minor progenitors of the coldest clusters (blue). All this material is rapidly stripped into the ICM once the satellites fall within $r_{200}$, diluting the ICM metallicity.

A more observationally-motivated way to look at this phenomenon is via the total stellar-mass-to-hot-gas-mass ratio, $M_{*,\tn{tot}}/M\sub{hot,tot}$. Panel B of Fig. \ref{fig:SatHists_NewModel} shows that $M_{*,\tn{tot}}/M\sub{hot,tot}$ positively correlates with $Z\sub{Fe}$, and Panel C shows that the hottest clusters typically have lower $M_{*,\tn{tot}}/M\sub{hot,tot}$ than the coldest clusters in \textsc{L-Galaxies} at $z=3.1$. Given that hotter clusters also have higher $M_{*,\tn{tot}}$, this is another indication that ICM metallicities were diluted in these systems due to excess accretion of pristine gas.

We therefore conclude that temperature-dependent dilution is a possible explanation for some of the residual $T$-$Z\sub{Fe}$ anti-correlation seen in the real cluster population.

\section{Summary \& Conclusions} \label{sec:Conclusions}
A homogenised dataset of \NumObsSystems groups and clusters (\NumObsMeasurements individual measurements) has been compiled, in order to study the $T$-$Z\sub{Fe}$ relation in the ICM with unprecedented accuracy. We correct for differences in aperture size, solar abundance, and cosmology among the samples used, and adopt $T$, $\rho\sub{gas}$, and $Z\sub{Fe}$ profiles that are adapted for each cluster individually (\S \ref{sec:Definitions_and_Derivations}).

This dataset is compared to model groups and clusters from the Munich semi-analytic model of galaxy evolution, \textsc{L-Galaxies}. This comparison allows us to (a) assess \textsc{L-Galaxies}'s ability to model massive systems, and (b) provide a physical explanation for those observed trends that the model does reproduce. Our main conclusions are as follows:

\begin{itemize}
\item Once homogenised, the scatter in the observed $T$-$Z\sub{Fe}$ relation for clusters is reduced significantly (\S \ref{sec:Obs_results}). The 1$\sigma$ dispersion in $Z\sub{Fe}$ of only \ObsTZscatter dex around a linear fit above $\Temp=0.25$ is comparable to the dispersion in the well-studied $M_{*}$-$Z\sub{O}$ relation for local, star-forming galaxies.

\item There is a slight anti-correlation between $T$ and $Z\sub{Fe}$ for clusters above $\Temp \sim 0.25$, with a slope of \ObsTZslope (\S \ref{sec:T-Z for clusters}). This anti-correlation could partly be explained by measurement biases, but is likely to also have a residual physical origin.

\item A possible explanation for this $T$-$Z\sub{Fe}$ anti-correlation is increased accretion of hydrogen by the most massive clusters, via stripping of infalling satellite systems. This is the cause of a weak anti-correlation in our galaxy evolution model, with a slope of \ModelTZslope (\S \ref{sec:Iron_in_model_clusters}).

\item The iron abundances observed in the hottest clusters are reasonably reproduced in \textsc{L-Galaxies} (\S \ref{sec:Iron_in_model_clusters}). This is true without requiring any changes to the rate of SNe-Ia, the IMF, or the metallicity of galactic winds assumed. This result is achieved while \textit{simultaneously} matching the chemical properties observed in the ISM of local star-forming galaxies, the G dwarfs in Milky-Way-type stellar discs, and the stellar populations of nearby ellipticals.

\item The iron abundance in intermediate-temperature clusters is under-estimated in our model by $\sim 0.2$ dex (\S \ref{sec:Iron_in_model_clusters}). This could partly be due to temperature-dependent biases in the way $Z\sub{Fe}$ is measured. However, it is possible that \textsc{L-Galaxies} is also not correctly modelling these systems. When treating this problem, we note that modifications to the GCE modelling to boost the enrichment of the ICM can also destroy the correspondence with smaller-scale systems.

\item The iron abundances and baryon fractions in galaxy groups are over-estimated in our model (\S \ref{sec:The baryon fraction} and \ref{sec:Iron_in_model_groups}). Allowing AGN feedback to remove metal-rich gas from galaxy groups is a viable solution to both these problems. Such re-modelling in \textsc{L-Galaxies} will be the focus of future work.

\item The model $z$-$Z\sub{Fe}$ relation for hot clusters is consistent with that seen in observations from $z=0.3$ to at least 1.27 (\S \ref{sec:Evo}). In \textsc{L-Galaxies}, the iron abundance in the ICM at $z=2$ is 66 per cent that seen at $z=0$. At $z=1$, it is 79 per cent, and at $z=0.5$, it is 85 per cent.
\end{itemize}

Despite the careful homogenisation process applied to our observational dataset in this work, our results are still conditional on a number of biases and uncertainties that have not been fully addressed. The significance of the instrumentation and spectral fitting codes used, temperature-dependent measurement biases, projected versus de-projected quantities, 1T versus multi-T spectral modelling, single-$\beta$ versus multi-$\beta$ gas density profiles, the variation in iron abundance profiles within the same class of system (\eg CCs), to name only a few, have not been investigated in detail here. However, it is promising that we are still able to obtain a small scatter in the $T$-$Z\sub{Fe}$ relation for clusters. This provides a clearer picture of the true iron abundances in groups and clusters than was possible before, with the promise of a still clearer picture once these additional effects are also accounted for.

In the case of our model results, it is important to note that, despite the improvement compared to previous theoretical studies, \textsc{L-Galaxies} still appears to be inadequate at modelling lower-temperature galaxy associations. This problem has also been highlighted by \citet{H15}. We caution that it is important that future modelling efforts, while attempting to resolve this issue, also account for the tight constraints provided by having to simultaneously reproduce the chemical properties of a wide range of galaxy systems.

\section*{Acknowledgments} \label{sec:Acknowledgements}
The authors would like to thank Jesper Rasmussen and Toru Sasaki for providing their data for comparison in this work. We would also like to thank Mike Anderson, Judith Croston, Jelle de Plaa, Dominique Eckert, Guinevere Kauffmann, Joe Mohr, Margherita Molaro, and Jesper Rasmussen for helpful discussions. RMY acknowledges the support of the Deutsche Forschungsgesellshaft (DFG) and through the Sofia Kovalevskaja Award to P. Schady from the Alexander von Humboldt Foundation of Germany. PAT (ORCID 0000-0001-6888-6483) acknowledges support from the Science and Technology Facilities Council (grant number ST/L000652/1). The work of BMBH was supported by Advanced Grant 246797 "GALFORMOD" from the European Research Council and by a Zwicky fellowship.

The authors contributed to this paper in the following ways: the observational data collection and recalibration, as well as the modifications to the \textsc{L-Galaxies} model, were undertaken by RMY. PAT provided the initial impetus for the project, routines to compute various DM halo properties, and advice and guidance on the interpretation of the observations. BMBH provided expertise and advice on the galaxy evolution modelling. The writing of the paper was undertaken by RMY, with proofreading by both PAT and BMBH.

\section*{Appendix A: Observational samples} \label{sec:Appendix A}
Here, we outline the ten different low-redshift observational studies we utilise in this work. These have been roughly separated into cluster and group samples below, although some of the cluster samples contain a few groups as we define them (\S \ref{sec:Definitions_and_Derivations}), and vice versa. Unless stated otherwise, quoted ICM temperatures and iron abundances are converted into $T_{500}$ and $\bar{Z}\sub{Fe,500}$ using our default radial profiles, as described in \S \ref{sec:Temperature estimation} and \ref{sec:Iron abundance estimation}.

The $kT_{500}$ - $\bar{Z}\sub{Fe,500}$ relations for each sample are shown in Fig. \ref{fig:kT-FeH_Obs_IndSamps_Clusters} for the cluster samples and Fig. \ref{fig:kT-FeH_Obs_IndSamps_Groups} for the group samples.

\renewcommand{\theequation}{A\arabic{equation}}
\setcounter{equation}{0}

\subsection*{A.1 Cluster samples}\label{sec:Cluster samples}
\subsubsection*{A.1.1 F98 \citep{F98}}
The oldest dataset utilised here is that of \citeauthor{F98} (1998, hereafter F98). Their study relied on \textit{ASCA} X-ray data for 40 galaxy groups and clusters, of which 34 could be used to obtain values of $T_{500}$ and $\bar{Z}\sub{Fe,500}$. Of these 34 objects (of which 6 are groups), 9 are identified as NCC. Mean temperatures (assumed to be emission weighted) and iron abundances were measured within a clustercentric annulus of inner radius $0.07\,h_{73}^{-1}$Mpc and outer radius $0.27\,h_{73}^{-1}$Mpc. Iron abundances are provided by F98 in their table 1, and are corrected from the \citet{AG89} \textit{photospheric} abundances originally used to those of \citet{GS98}.

\begin{figure*}
\centering
\includegraphics[width=0.95\textwidth]{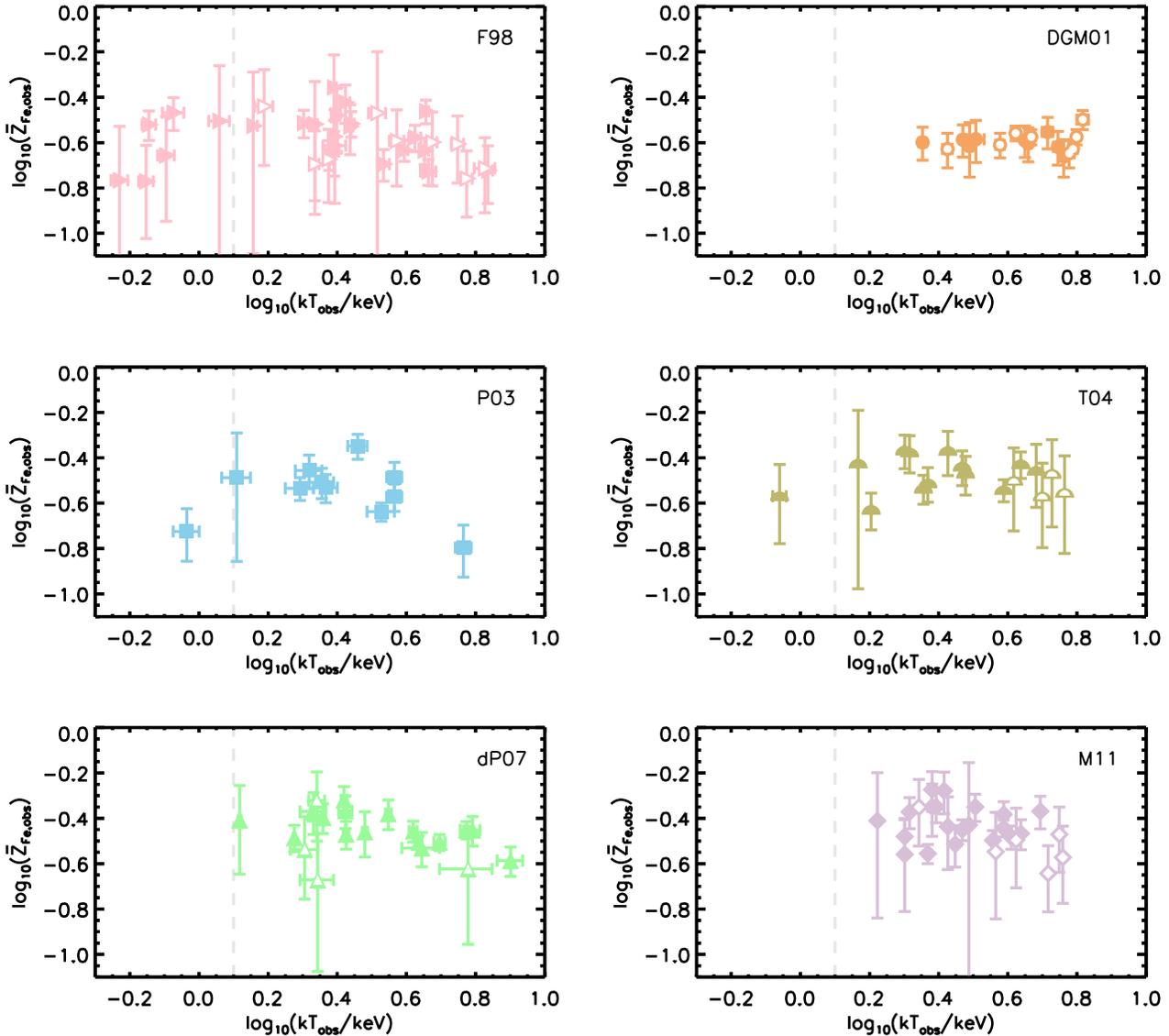}
\caption{The $kT_{500}$-$\bar{Z}\sub{Fe,500}$ relation for each of the cluster samples in our dataset. Filled symbols indicate CC systems and open symbols indicate NCC systems. The grey vertical line separates groups and clusters.}
\label{fig:kT-FeH_Obs_IndSamps_Clusters}
\end{figure*}

\subsubsection*{A.1.2 DGM01 \citep{DGM01}}
\citeauthor{DGM01} (2001, hereafter DGM01) observed 17 galaxy clusters with \textit{BeppoSAX}, obtaining projected iron abundance profiles out to a maximum radius between 8 and 20 arcminutes (equivalent to between $0.4$ and $1.2\,h_{73}^{-1}$Mpc, or 0.2 and $0.5r_{180}$), the largest outer radii of any survey considered here. All 17 of these clusters (9 CC and 8 NCC) could be used to obtain $T_{500}$ and $\bar{Z}\sub{Fe,500}$. Values of $\bar{T}\sub{ew}$ from table 2 of the companion paper by \citet{DGM02} were used to obtain $T_{500}$.

$\bar{Z}\sub{Fe,500}$ was obtained from the average iron abundances given in their table 2, rescaling from the \textit{meteoric} abundances of \citet{AG89} that were originally used to those of \citet{GS98}. DGM01 reported their own typical iron abundance profile for their CC clusters, given by our Eqn. \ref{eqn:Z_profile_DGM01}, with values of $x_{\tn{c}}=0.04$ and $\alpha=0.18$. Given that their average iron abundances are obtained from fitting a constant to their radial measurements, we choose to use their profile for each of their CC clusters, rather than our default one. We note that the difference in the value of $\bar{Z}\sub{Fe,500}$ obtained is less than 0.02 dex, due to the large aperture size used. DGM01 conclude that the iron abundance profile in NCC clusters is effectively flat. Later works have instead argued that NCC clusters do have negative gradients, similar to or slightly shallower than those found in CC clusters (\eg \citealt{Ta04,Sa09,M11}). Here, we take a conservative approach, by assuming a flat gradient for the DGM01 NCC clusters, as DGM01 suggested. Assuming a steeper gradient would lower the final estimated value of $\bar{Z}\sub{Fe,500}$ by 0.05-0.10 dex.

A3627 is classified as having a CC, based on the temperature profile measured by \citet{DGM02}. The merging system A3266 has a measured CCT of $7.51\,h_{73}^{-1/2}$Gyr and therefore has a weak CC under the \citet{H10} definition. However, given that A3266 also has a classical mass deposition rate consistent with zero \citep{Wh97,Pe98,H10}, a central-to-virial temperature ratio greater than one \citep{H10}, and that the low-entropy gas in its core region is more likely to be stripped from an infalling sub-cluster than to be a signature of a cool core \citep{Fi06a}, we choose to define this cluster as NCC in this work.

\subsubsection*{A.1.3 P03 \citep{P03}}
\citeauthor{P03} (2003, hereafter P03) observed 14 CC systems with \textit{XMM-Newton}, to obtain mean iron abundances, as well as O, Ne, Mg, and Si abundances, which we rescale from the \textit{meteoric} abundances of \citet{AG89} originally used to those of \citet{GS98}. Enough data to derive $T_{500}$ and $\bar{Z}\sub{Fe,500}$ estimates are available for 11 of these objects. We assume that the measurements of temperature and abundance are representative of the cluster core, whose physical radius can be obtained from $r\sub{core}$/arcsec in their table 3 and the angular scale from their table 2.

We further assume the quoted ``ambient'' or ``upper'' temperature (from their table 5) is the peak temperature of the X-ray-emitting gas in the cluster, and correct this to $\bar{T}\sub{mw}$ using the factor of 1.21 provided by \citeauthor{V06} (2006, eqn. 9).

Only one of the P03 objects (NGC533) has $T_{500} < 0.1$ keV and so is classified as a group in this work. This means that the correction to obtain $\bar{Z}\sub{Fe,500}$ for NGC533 is larger than for the rest of the P03 sample, as it utilises our steeper group profile (\S \ref{sec:Iron abundance estimation}). The classification of CC for A1835 was taken from the analysis of \citet{Sc01} using \textit{Chandra} data. They found a steep drop in ICM temperature at low radii, indicative of a cool core.

The unexpectedly high iron abundance of $\tn{log}(\bar{Z}\sub{Fe,500}) = -0.35$ obtained for A496 from the P03 sample data (see Fig \ref{fig:FeH_comp_Obs}) is likely due to our assumption that their average abundances are representative of the entire cluster core. P03 measured the core radius of A496 as $255\,h_{73}^{-1}$kpc, which is significantly larger than the $20.5\,h_{73}^{-1}$kpc calculated by \citet{RB02} and also larger than the 2'-diameter square aperture ($80.5\,h_{73}^{-1}$kpc for A496) within which P03 selected photons (see their section 4). We can therefore assume that our derived $\bar{Z}\sub{Fe,500}$ is over-estimated in this particular case.

\subsubsection*{A.1.4 T04 \citep{Ta04}}
\citeauthor{Ta04} (2004, hereafter T04) used \textit{XMM-Newton} spectra of 19 nearby systems to obtain mean iron abundances, as well as O, Si, and S abundances. We rescale these from the \textit{photometric} abundances of \citet{AG89} originally used to those of \citet{GS98}. 17 of these objects (1 group, 16 clusters), of which 4 are NCC, are suitable for obtaining $T_{500}$ and $\bar{Z}\sub{Fe,500}$ estimates. As advised by T04, MKW9 is not included in our analysis, due to the very high uncertainty in its iron abundance. As mentioned in \S A.1.3, cluster A1835 is classified as containing a CC. Mean (emission-weighted) temperatures outside the cool region are taken from their table 1 and used to obtain $T_{500}$. No errors are quoted by T04 for their mean temperature measurements.

In order to retain as many of the T04 sample as possible, abundances from the intermediate annulus of $0.07-0.27\,h_{73}^{-1}$Mpc (\ie from $0.03-0.1r_{180}$ to $0.1-0.4r_{180}$) are used (their table 4), rather than the overall mean abundances provided for a sub-set of their objects. As always, we account for the inner and outer observed radii when calculating $\bar{Z}\sub{Fe,500}$. T04 note that they see no clear difference between the iron abundance gradients in their CC and NCC clusters (their section 5.2), with both showing central enhancements. They cite the improved spatial resolution of \textit{XMM-Newton} as the reason for this finding, compared to previous conclusions (\eg \citealt{DGM01}). Here, we again take a conservative approach and assume our default CC profile for CC clusters and our shallower default NCC profile for NCC clusters (\S \ref{sec:Iron abundance estimation}). If we were to assume a steeper profile for NCC clusters, it would serve only to further reduce the estimated value of $\bar{Z}\sub{Fe,500}$ by $\sim 0.05$ dex for these objects (\S A.1.7).

As noted in \S \ref{sec:Definitions_and_Derivations}, measurements of $r\sub{c}$ and $\beta$ compiled by \citet{RB02} are used for all our cluster samples where possible. One exception to this rule is A399, for which an uncertain core radius ($\sigma(r\sub{c})/r\sub{c}=0.29$) is quoted by \citet{RB02}. In this case, we rely on the more recent measurements of \citet{SP04}, who measured $r\sub{c}=148.9\,h^{-1}_{73}$kpc, $\sigma(r\sub{c})=0.5\,h^{-1}_{73}$kpc, $\beta=0.498$ and $\sigma(\beta)=0.001$.

\subsubsection*{A.1.5 dP07 \citep{dP07}}
\citeauthor{dP07} (2007, hereafter dP07) also used \textit{XMM-Newton} archive data of 22 nearby clusters to obtain mean temperatures and iron abundances, as well as abundances of Si, S, Ar, Ca, and Ni. All but one of these measurements could be used for our analysis, or which 4 are NCC. The extraction radius within which measurements were taken was $0.2\,r_{500}$.

As the dP07 extraction radius does not go all the way out to $r_{500}$, we use the maximum ICM temperatures quoted in their table 2 to obtain $T_{500}$, rather than the mean temperatures from the inner regions only. As the \citet{V06} temperature profile we use itself peaks at around $0.2\,r_{500}$, we can assume that the maximum temperature measured by dP07 is the true peak temperature within the whole cluster. Mean iron abundances are taken from their table A.1, and converted from the \textit{proto-solar} abundances of \citet{L03} originally used to those of \citet{GS98}.

A3530 was classified as an NCC cluster, based on the high cooling time and zero mass deposition rate measured by \citet{Ch07}. And A3560 was classified as having an NCC, based on the flat temperature profile within $\sim 0.4\,h_{73}^{-1}$Mpc measured by \citet{Ba02} using \textit{BeppoSAX} data.

\subsubsection*{A.1.6 M11 \citep{M11}}
\citeauthor{M11} (2011, hereafter M11) analysed 28 galaxy clusters observed by \textit{XMM-Newton}, of which 26 could be used here (6 being classified as NCC clusters). All objects have $T_{500} > 0.1$ keV.

We note here that, although M11 have classified cluster A3558 as not containing a central cD galaxy (and therefore unlikely to have a cool core), the mass deposition rate measured by \citet{Pe98} for this object is non-zero. Therefore, we classify A3558 as a CC cluster in this work, in line with the CC classification for this object by dP07. M11 provides mean (emission-weighted) temperatures within an annulus of $0.06-0.3\,r_{180}$. As mentioned in \S A.1.2, cluster A3627 is classified as a NCC cluster.

Iron abundances within the $0.03-0.06\,r_{180}$ annulus are taken from their table 2 and converted from the \textit{photospheric} abundances of \citet{L03} originally used to those of \citet{GS98}. The cluster iron abundance profiles we use to obtain $\bar{Z}\sub{Fe,500}$ are fit to the stacked M11 data, so we can expect them to be a particularly good representation of the typical iron abundance gradients in this sample.

\subsubsection*{A.1.7 Cluster sample $T$-$Z\sub{\textit{Fe}}$ relations}
Fig. \ref{fig:kT-FeH_Obs_IndSamps_Clusters} shows the $kT_{500}$ - $\bar{Z}\sub{Fe,500}$ relations for each of the cluster samples described above.

The three samples on the left of Fig. \ref{fig:kT-FeH_Obs_IndSamps_Clusters} exhibit clear negative correlations between $T$ and $Z\sub{Fe}$. Simple linear fits to each of these three samples above $\Temp=0.25$ yield slopes ranging from -0.23 (dP07 sample) to -0.39 (P03 sample). These are comparable to the slope of \ObsTZslope obtained for the complete dataset. The three samples on the right of Fig. \ref{fig:kT-FeH_Obs_IndSamps_Clusters} exhibit flatter relations however, with slopes ranging from -0.03 (DGM01 sample) to -0.17 (T04 sample).

We note that the T04 relation would become steeper if we were to assume the steeper iron abundance profiles they measure for their NCC clusters, as discussed in \S A.1.4. The very flat $T$-$Z\sub{Fe}$ relation for the DGM01 sample originates from the original mean, emission-weighted iron abundances, which also have no clear correlation with $T_{500}$. However, \citet{DG04} have found a slight negative slope for the same set of clusters when considering the additional information on the ICM gas mass provided by \citet{E02} to obtain $M\sub{Fe}$/$M\sub{gas}$ within $r_{2500}$ and $r_{1000}$ (see \eg their fig. 8).

We therefore conclude that a slight negative slope is a common feature of the $T$-$Z\sub{Fe}$ relation, which at least warrants further discussion (\S \ref{sec:T-Z for clusters}).

\subsection*{A.2 Group samples}\label{sec:Group samples}
\subsubsection*{A.2.1 M05 \citep{Mah05}}
\citeauthor{Mah05} (2005, hereafter M05) studied \textit{XMM-Newton} spectra of 7 galaxy groups and 1 galaxy cluster (A2634), originally from the \textit{ROSAT} RASSCALS catalogue. All 8 can be used in our analysis, of which 1 is NCC. M05 measured (emission-weighted) mean ICM temperatures between $0.1\,r_{500}$ and $0.5\,r_{500}$.

Iron abundances were measured within the same annulus, and are corrected from the \textit{photospheric} abundances of \citet{AG89} originally used to those of \citet{GS98}. We note that group NRGb184 exhibits a particularly low iron abundance within the measured annulus of only $\tn{log}(\bar{Z}_{0.1-0.5r_{500}}) = -1.0$. A similarly low value was also found by \citet{J11}, who attributed it to member galaxies ejecting an usually low amount of metals into the ICM for their stellar mass. However, \citet{J11} also determined that the iron abundance gradient in NRGb184 is effectively flat (their fig. 3), in contrast to typical CC groups. Based on this information, we make an exception for NRGb184, and assume a flat abundance gradient when calculating $\bar{Z}\sub{Fe,500}$ rather than the default \citet{RP07} profile used for other groups.

Values of the gas-density slope, $\beta$, were not obtainable for four of the M05 groups, A194, NGC3411, NRGb184, and NGC5098. In these cases, we estimate $\beta$ from the ICM temperature, as is done for our model groups and clusters, via the $\bar{T}\sub{ew}$-$\beta$ relation provided by \citet{Sa03} given by Eqn. \ref{eqn:T-beta}. Similarly, we assume $r\sub{c}$ is equal to the scale length, $a=r_{200}/c$, for all M05 objects, except for the cluster A2634, for which a measurement of $r\sub{c}$ is provided by \citet{RB02}.

\subsubsection*{A.2.2 F06 \citep{Fi06}}
\citeauthor{Fi06} (2006, hereafter F06) studied archival data of 11 galaxy groups from the \textit{XMM-Newton} Group Survey (2dXGS). We choose to take values of $\bar{T}\sub{mw}$ and $\bar{Z}\sub{Fe}$ from their $0.1-0.5r_{500}$ annulus, to be in better correspondence with the profile fits we are using here. This leaves 6 groups available for our analysis, of which 1 is NCC. NGC4168 is excluded, as F06 estimate zero metallicity within this annulus. NGC4168 is, in fact, another object which \citet{J11} determine to be unusually inefficient at polluting the ICM with metals for its stellar mass (\S A.2.1). Iron abundances for the F06 sample were corrected from the \textit{photospheric} abundances of \citet{AG89} originally used to those of \citet{GS98}.

For one group, NGC4261, values for $\beta$ and $r\sub{c}$ could not be obtained. Therefore, as for those cases in the M05 sample, $\beta$ is inferred from the ICM temperature and we assume $r\sub{c}=a$.

\begin{figure}
\centering
\includegraphics[width=0.45\textwidth]{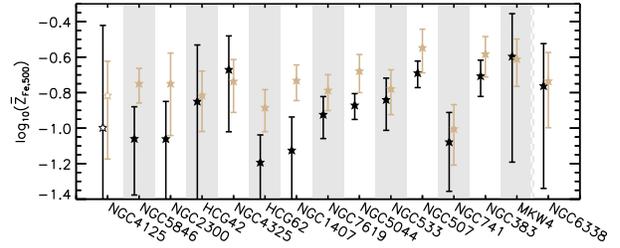} \\
\caption{A comparison of the mass-weighted iron abundances obtained from the iron masses provided by RP09 and our Eqn. \ref{eqn:RP09_abundances} (brown), with those obtained by correcting the emission-weighted iron abundances provided by \citet{RP07} using our default group $Z\sub{Fe}$ profile (\S \ref{sec:Iron abundance estimation}) (black).}
\label{fig:RP07_vs_RP09_ZFe}
\end{figure}

\begin{figure}
\centering
\includegraphics[width=0.45\textwidth]{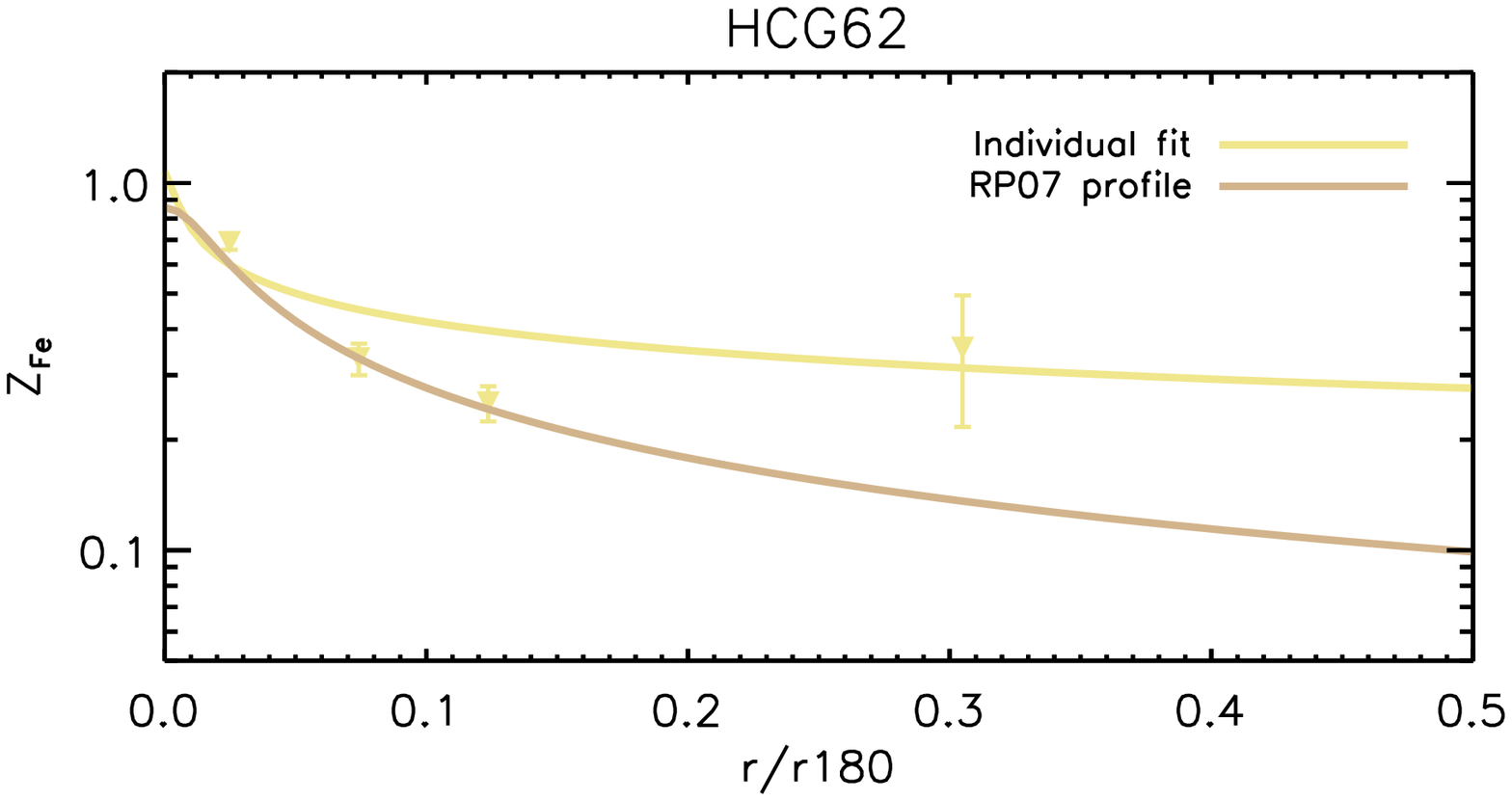} \\
\vspace{0.1in}
\includegraphics[width=0.45\textwidth]{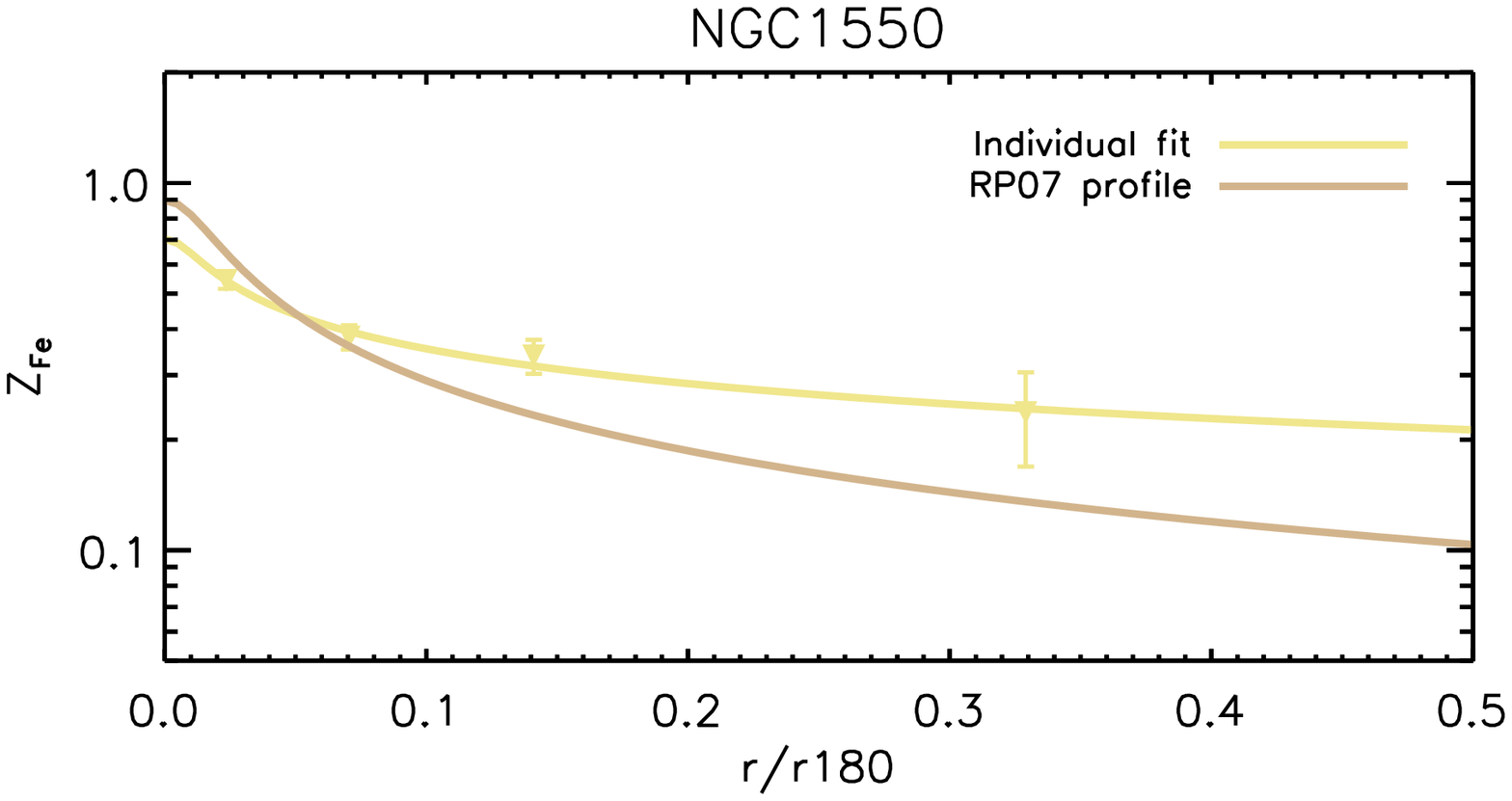} \\
\vspace{0.1in}
\includegraphics[width=0.45\textwidth]{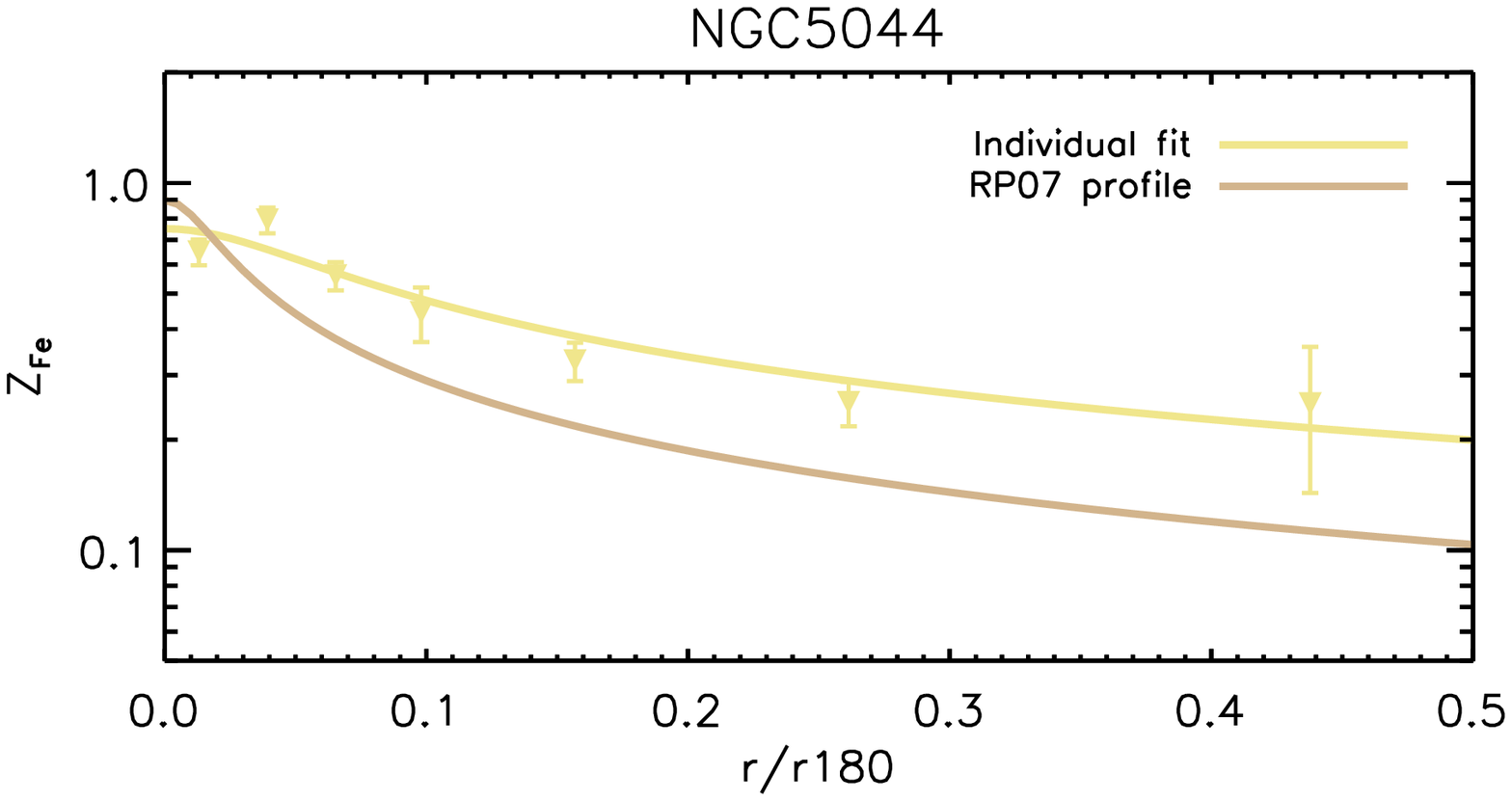} \\
\caption{Iron abundance profiles for three groups from the S14 sample: HCG62, NGC1550, and NGC5044. Yellow points indicate radial data measured by S14. Yellow lines are fits to the data using Eqn. \ref{eqn:Z_profile_DGM01} and treating $x\sub{c}$ and $\alpha$ as free parameters. Brown lines are fits to the data using the same equation and fixing $x\sub{c}$ and $\alpha$ to their default values for groups, as given in Table \ref{tab:Z_profile_params}. Profiles have been rescaled to $r_{180}$ in this figure, assuming $r_{500} = 0.64\,r_{180}$.}
\label{fig:S14 Z profiles}
\end{figure}

\begin{figure*}
\centering
\includegraphics[width=0.95\textwidth]{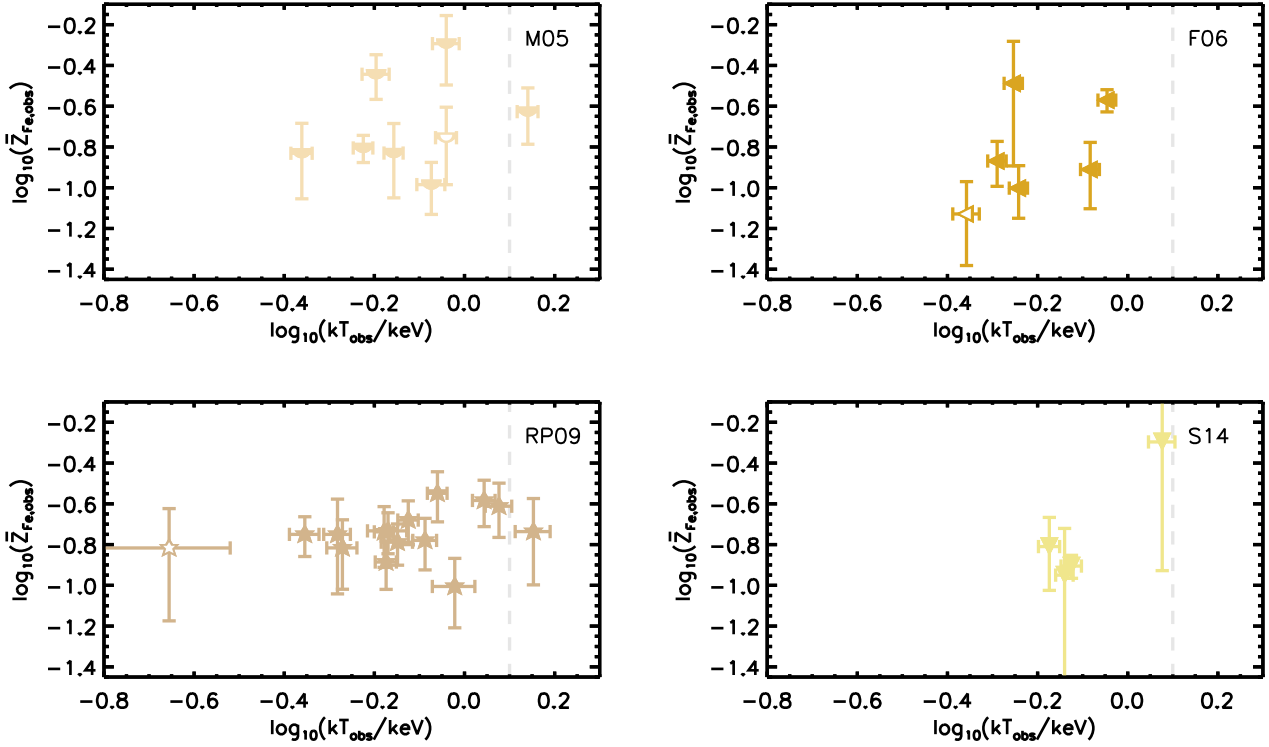}
\caption{The $kT_{500}$-$\bar{Z}\sub{Fe,500}$ relation for each of the group samples in our dataset. Filled symbols indicate CC systems and open symbols indicate NCC systems. The grey vertical line separates groups and clusters.}
\label{fig:kT-FeH_Obs_IndSamps_Groups}
\end{figure*}

\subsubsection*{A.2.3 RP09 \citep{RP09}}
\citeauthor{RP09} (2009, hereafter RP09) studied \textit{Chandra} data of 14 groups (1 NCC) and 1 cluster to obtain iron, silicon, and total gas \textit{masses} within $r_{500}$ (their table 1). This was achieved by assuming a $\beta$-model gas density profile and the abundance gradient measurements from their previous work, \citet{RP07}. We use these masses to directly obtain ICM iron abundances as follows;

\begin{equation}\label{eqn:RP09_abundances}
	\tn{log}(\bar{Z}\sub{Fe,500}) = \tn{log}\left(\frac{M\sub{Fe,500}/A\sub{Fe}}{f\sub{H}\,M\sub{gas,500}/A\sub{H}}\right) - \tn{log}\left(\frac{N\sub{Fe,\astrosun}}{N\sub{H,\astrosun}}\right)\;\;,
\end{equation}
where $A\sub{Fe}$ and $A\sub{H}$ are the atomic weights of iron and hydrogen respectively, and $f\sub{H}$ is the fraction of the total ICM gas mass expected to be hydrogen. This fraction is given by $f\sub{H} = [1-Y\asun-(Z\asun\cdot{}\bar{Z}_{0.1-0.3r_{500}})]$, where we assume the solar fraction of helium, $Y\asun$, and use the average metallicity measured by \citet{RP07} between 0.1 and $0.3\,r_{500}$.

RP09 adopted the same solar abundances from \citet{GS98} as we have in this work, so no further correction is required here. Mean, emission-weighted temperatures were also measured within a $0.1 - 0.3\,r_{500}$ annulus.

These RP09 mass-weighted iron abundances utilise radial profiles that have been individually fit to each object, and so provide a good benchmark with which to compare the iron abundances obtained using our default group profile. In Fig. \ref{fig:RP07_vs_RP09_ZFe}, we directly compare the iron abundances derived from the RP09 iron masses (brown), with those obtained using our homogenisation process and the emission-weighted iron abundances provided by \citet{RP07} for the same systems (black). We can see that the values are very similar in most cases, further indicating that our homogenisation process is working well at producing realistic mass-weighted iron abundances. We note that those objects which show a larger discrepancy in Fig. \ref{fig:RP07_vs_RP09_ZFe} (\ie NGC5846, NGC2300, HCG62, and NGC1407) all have \textit{lower} emission-weighted metallicities (from \citealt{RP07}) than mass-weighted metallicities (obtained from the RP09 iron masses). Our homogenisation process is unable to reproduce such surprising cases, as the emission-weighted measure will always be larger than the mass-weighted measure.

As noted in \S \ref{sec:Definitions_and_Derivations}, measurements of $r\sub{c}$ and $\beta$ compiled by RP09 an \citet{M03} are used for all our group samples where possible. An exception to this rule is NGC533, for which a particularly high error on $r\sub{c}$ is quoted by RP09 [$r\sub{c}=2.2\,h^{-1}_{73}$kpc and $\sigma(r\sub{c})=1.7\,h^{-1}_{73}$kpc], and only an upper limit of $r\sub{c}<2.15\,h^{-1}_{73}$kpc is provided by \citet{M03}. In this case, we rely on the more recent measurements of \citet{Gu12}, which give $r\sub{c}=1.64\,h^{-1}_{73}$kpc and $\sigma(r\sub{c})=0.10\,h^{-1}_{73}$kpc.

\subsubsection*{A.2.4 S14 \citep{S14}}
\citeauthor{S14} (2014, hereafter S14) analysed \textit{Suzaku} data of 4 galaxy groups (all CC) which had been observed out to $\sim 0.5\,r_{180}$. For two of these groups, NGC1550 and NGC5044, we individually fit iron abundance profiles of the form given by Eqn. \ref{eqn:Z_profile_DGM01}, using radial abundance data kindly provided by T. Sasaki (priv. comm.), which is calculated assuming a two-temperature (2T) model. We allow all three parameters in the $Z\sub{Fe}$ profile (namely, $Z\sub{Fe,0}$, $x\sub{c}$, and $\alpha$) to be simultaneously fit. A comparison between these individual fits, our default group profile fits (\ie with our default values for $x\sub{c}$ and $\alpha$), and the original S14 data are shown in Fig. \ref{fig:S14 Z profiles}.

For HCG62 (top panel of Fig. \ref{fig:S14 Z profiles}), we choose to rely on our default group profile fit (brown line), as the individual fit for this cluster is strongly biased by the outermost radial measurement, which is surprisingly high. Our default group profile, normalised to $Z\sub{Fe,0}$ as described in \S \ref{sec:Iron abundance estimation}, is a much closer match to the other three data points, and so is preferred here. We also rely on our default group profile for MKW4, as not enough radial information is provided to individually constrain $x\sub{c}$ and $\alpha$ for this object.

All abundances were corrected from the \textit{photospheric} abundances of \citet{L03} originally used to those of \citet{GS98}. Mean, emission-weighted temperatures are taken from the RP09 sample, which also contains these 4 objects. Redshifts and structural parameters are also available for all of the S14 groups from the standard catalogues mentioned in \S \ref{sec:Definitions_and_Derivations}.

\subsubsection*{A.2.5 Group sample $T$-$Z\sub{\textit{Fe}}$ relations}
Fig. \ref{fig:kT-FeH_Obs_IndSamps_Groups} shows the $kT_{500}$ - $\bar{Z}\sub{Fe,500}$ relations for each of the group samples described above.

\begin{table*}
\centering
\begin{tabular}{p{1.35in}ccccccc}
\hline \hline
$^{1}$\textbf{Name} & $^{2}$\textbf{Class} & $^{3}$\textbf{Type} & $^{4}$\textbf{Sample} & $^{5}$$\bs{kT_{500}}$ & $^{6}$$\bs{\sigma(kT_{500})}$ & $^{7}$$\bs{\bar{Z}\sub{\textbf{Fe,500}}}$ & $^{8}$$\bs{\sigma(\bar{Z}\sub{\textbf{Fe,500}})}$ \\
\hline
2A 0335+096 \dotfill & Cl & CC & F98 & 2.01 & $\pm$0.05 & 0.31 & $\pm0.04$ \\
 & & & DGM01 & 2.26 & $\pm$0.05 & 0.25 & $\pm0.04$ \\
 & & & P03 & 1.96 & $\pm$0.18 & 0.29 & $\pm0.03$ \\
 & & & T04 & 2.00 & -- & 0.43 & $\pm0.07$ \\
 & & & dP07 & 2.14 & $\pm$0.01 & 0.43 & $\pm0.04$ \\
 & \vspace{-0.095in} \\
A85 \dotfill & Cl & CC & F98 & 4.21 & $\pm$0.17 & 0.26 & $\pm0.04$ \\
 & & & DGM01 & 4.56 & $\pm$0.10 & 0.25 & $\pm0.04$ \\
 & & & dP07 & 4.17 & $\pm$0.11 & 0.35 & $\pm0.04$ \\
 & & & M11 & 3.87 & -- & 0.41 & $\pm0.06$ \\
 & \vspace{-0.095in} \\
A119 \dotfill & Cl & NCC & F98 & 3.73 & $\pm$0.18 & 0.26 & $\pm0.09$ \\
 & & & DGM01 & 3.78 & $\pm$0.11 & 0.25 & $\pm0.03$ \\
 & \vspace{-0.095in} \\
A133 \dotfill & Cl & CC & dP07 & 2.63 & $\pm$0.05 & 0.48 & $\pm0.07$ \\
 & \vspace{-0.095in} \\
A194 \dotfill & Gr & NCC & M05 & 0.91 & $\pm$0.05 & 0.18 & $\pm0.07$ \\
 & \vspace{-0.095in} \\
A262 \dotfill & Cl & CC & F98 & 1.44 & $\pm$0.04 & 0.30 & $\pm0.22$ \\
 & & & P03 & 1.29 & $\pm$0.12 & 0.33 & $\pm0.19$ \\
 & & & T04 & 1.47 & -- & 0.37 & $\pm0.27$ \\
 & & & M11 & 1.67 & -- & 0.39 & $\pm0.24$ \\
 & \vspace{-0.095in} \\
A399 \dotfill & Cl & NCC & T04 & 4.14 & -- & 0.31 & $\pm0.13$ \\
 & \vspace{-0.095in} \\
A400 \dotfill & Cl & NCC & F98 & 1.54 & $\pm$0.09 & 0.36 & $\pm0.16$ \\
 & \vspace{-0.095in} \\
A426 (Perseus) \dotfill & Cl & CC & F98 & 4.53 & $\pm$0.08 & 0.35 & $\pm0.04$ \\
 & & & DGM01 & 4.46 & $\pm$0.05 & 0.26 & $\pm0.04$ \\
 & & & T04 & 4.34 & -- & 0.37 & $\pm0.05$ \\
 & & & M11 & 4.07 & -- & 0.33 & $\pm0.04$ \\
 & \vspace{-0.095in} \\
A478 \dotfill & Cl & CC & F98 & 4.61 & $\pm$0.23 & 0.19 & $\pm0.02$ \\
 & \vspace{-0.095in} \\
A496 \dotfill & Cl & CC & F98 & 2.76 & $\pm$0.05 & 0.30 & $\pm0.04$ \\
 & & & DGM01 & 2.95 & $\pm$0.05 & 0.26 & $\pm0.04$ \\
 & & & P03 & 2.88 & $\pm$0.18 & 0.45 & $\pm0.06$ \\
 & & & T04 & 2.94 & -- & 0.36 & $\pm0.06$ \\
 & & & M11 & 2.94 & -- & 0.35 & $\pm0.04$ \\
 & \vspace{-0.095in} \\
A539 \dotfill & Cl & NCC & F98 & 2.16 & $\pm$0.06 & 0.20 & $\pm0.08$ \\
 & \vspace{-0.095in} \\
A754 \dotfill & Cl & NCC & DGM01 & 6.29 & $\pm$0.11 & 0.27 & $\pm0.02$ \\
 & & & T04 & 5.34 & -- & 0.34 & $\pm0.14$ \\
 & & & M11 & 5.74 & -- & 0.27 & $\pm0.10$ \\
 & \vspace{-0.095in} \\
A1060 \dotfill & Cl & CC & F98 & 2.16 & $\pm$0.04 & 0.30 & $\pm0.16$ \\
 & & & M11 & 2.00 & -- & 0.28 & $\pm0.12$ \\
 & \vspace{-0.095in} \\
A1367 \dotfill & Cl & NCC & F98 & 2.37 & $\pm$0.05 & 0.21 & $\pm0.08$ \\
 & & & M11 & 2.20 & -- & 0.45 & $\pm0.15$ \\
 & \vspace{-0.095in} \\
A1644 \dotfill & Cl & CC & M11 & 3.07 & -- & 0.37 & $\pm0.33$ \\
 & \vspace{-0.095in} \\
A1651 \dotfill & Cl & CC & dP07 & 4.41 & $\pm$0.55 & 0.29 & $\pm0.05$ \\
 & \vspace{-0.095in} \\
A1656 (Coma) \dotfill & Cl & NCC & F98 & 5.60 & $\pm$0.23 & 0.25 & $\pm0.08$ \\
 & & & DGM01 & 6.14 & $\pm$0.09 & 0.24 & $\pm0.01$ \\
 & & & T04 & 5.01 & -- & 0.27 & $\pm0.11$ \\
 & & & M11 & 5.21 & -- & 0.23 & $\pm0.07$ \\
 & \vspace{-0.095in} \\
A1689 \dotfill & Cl & CC & dP07 & 7.96 & $\pm$0.67 & 0.26 & $\pm0.04$ \\
 & \vspace{-0.095in} \\
A1775 \dotfill & Cl & NCC & dP07 & 2.19 & $\pm$0.12 & 0.48 & $\pm0.16$ \\
 & \vspace{-0.095in} \\
A1795 \dotfill & Cl & CC & F98 & 3.93 & $\pm$0.09 & 0.23 & $\pm0.03$ \\
 & & & P03 & 3.37 & $\pm$0.31 & 0.23 & $\pm0.02$ \\
 & & & T04 & 3.87 & -- & 0.29 & $\pm0.03$ \\
 & & & dP07 & 4.32 & $\pm$0.08 & 0.31 & $\pm0.02$ \\
 & & & M11 & 3.87 & -- & 0.36 & $\pm0.03$ \\
 & \vspace{-0.095in} \\
A1835 \dotfill & Cl & CC & P03 & 5.82 & $\pm$0.31 & 0.16 & $\pm0.04$ \\
 & & & T04 & 4.81 & -- & 0.35 & $\pm0.11$ \\
 & \vspace{-0.095in} \\
A2029 \dotfill & Cl & CC & DGM01 & 5.19 & $\pm$0.19 & 0.28 & $\pm0.04$ \\
 & & & dP07 & 5.94 & $\pm$0.25 & 0.35 & $\pm0.03$ \\
 & & & M11 & 4.94 & -- & 0.43 & $\pm0.07$ \\
\hline
\end{tabular}
\caption{Final observational dataset\ldots{}}
\label{tab:Final_data0}
\end{table*}
 
\begin{table*}
\centering
\begin{tabular}{p{1.35in}ccccccc}
\hline \hline
$^{1}$\textbf{Name} & $^{2}$\textbf{Class} & $^{3}$\textbf{Type} & $^{4}$\textbf{Sample} & $^{5}$$\bs{kT_{500}}$ & $^{6}$$\bs{\sigma(kT_{500})}$ & $^{7}$$\bs{\bar{Z}\sub{\textbf{Fe,500}}}$ & $^{8}$$\bs{\sigma(\bar{Z}\sub{\textbf{Fe,500}})}$ \\
\hline
A2052 \dotfill & Cl & CC & P03 & 2.08 & $\pm$0.18 & 0.35 & $\pm0.06$ \\
 & & & T04 & 2.07 & -- & 0.42 & $\pm0.08$ \\
 & & & dP07 & 2.28 & $\pm$0.02 & 0.40 & $\pm0.06$ \\
 & & & M11 & 2.07 & -- & 0.42 & $\pm0.07$ \\
 & \vspace{-0.095in} \\
A2063 \dotfill & Cl & CC & F98 & 2.46 & $\pm$0.07 & 0.23 & $\pm0.05$ \\
 & & & M11 & 2.60 & -- & 0.53 & $\pm0.11$ \\
 & \vspace{-0.095in} \\
A2142 \dotfill & Cl & CC & DGM01 & 5.78 & $\pm$0.15 & 0.22 & $\pm0.04$ \\
 & \vspace{-0.095in} \\
A2147 \dotfill & Cl & NCC & F98 & 3.28 & $\pm$0.19 & 0.34 & $\pm0.29$ \\
 & \vspace{-0.095in} \\
A2199 \dotfill & Cl & CC & F98 & 2.74 & $\pm$0.05 & 0.30 & $\pm0.07$ \\
 & & & DGM01 & 3.08 & $\pm$0.07 & 0.24 & $\pm0.07$ \\
 & & & dP07 & 3.02 & $\pm$0.05 & 0.35 & $\pm0.08$ \\
 & & & M11 & 2.80 & -- & 0.31 & $\pm0.06$ \\
 & \vspace{-0.095in} \\
A2204 \dotfill & Cl & CC & dP07 & 6.19 & $\pm$0.31 & 0.35 & $\pm0.05$ \\
 & \vspace{-0.095in} \\
A2256 \dotfill & Cl & NCC & F98 & 4.73 & $\pm$0.15 & 0.25 & $\pm0.09$ \\
 & & & DGM01 & 4.65 & $\pm$0.08 & 0.27 & $\pm0.02$ \\
 & & & M11 & 4.21 & -- & 0.32 & $\pm0.12$ \\
 & \vspace{-0.095in} \\
A2319 \dotfill & Cl & NCC & F98 & 5.94 & $\pm$0.23 & 0.17 & $\pm0.06$ \\
 & & & DGM01 & 6.56 & $\pm$0.25 & 0.32 & $\pm0.03$ \\
 & \vspace{-0.095in} \\
A2589 \dotfill & Cl & CC & dP07 & 2.14 & $\pm$0.18 & 0.43 & $\pm0.09$ \\
 & & & M11 & 2.40 & -- & 0.53 & $\pm0.11$ \\
 & \vspace{-0.095in} \\
A2634 (NGC7720) \dotfill & Cl & CC & F98 & 2.47 & $\pm$0.19 & 0.24 & $\pm0.11$ \\
 & & & M05 & 1.38 & $\pm$0.07 & 0.24 & $\pm0.07$ \\
 & \vspace{-0.095in} \\
A3112 \dotfill & Cl & CC & T04 & 3.00 & -- & 0.34 & $\pm0.07$ \\
 & & & dP07 & 3.53 & $\pm$0.07 & 0.42 & $\pm0.06$ \\
 & \vspace{-0.095in} \\
A3266 \dotfill & Cl & NCC & DGM01 & 5.99 & $\pm$0.20 & 0.23 & $\pm0.03$ \\
 & & & T04 & 5.81 & -- & 0.28 & $\pm0.13$ \\
 & & & M11 & 5.61 & -- & 0.34 & $\pm0.11$ \\
 & \vspace{-0.095in} \\
A3376 \dotfill & Cl & NCC & DGM01 & 2.66 & $\pm$0.09 & 0.24 & $\pm0.04$ \\
 & \vspace{-0.095in} \\
A3526 (Centaurus) \dotfill & Cl & CC & F98 & 2.46 & $\pm$0.04 & 0.44 & $\pm0.17$ \\
 & & & M11 & 2.67 & -- & 0.37 & $\pm0.13$ \\
 & \vspace{-0.095in} \\
A3530 \dotfill & Cl & NCC & dP07 & 2.21 & $\pm$0.25 & 0.21 & $\pm0.13$ \\
 & \vspace{-0.095in} \\
A3558 \dotfill & Cl & CC & F98 & 3.42 & $\pm$0.13 & 0.20 & $\pm0.03$ \\
 & & & dP07 & 4.96 & $\pm$0.18 & 0.31 & $\pm0.03$ \\
 & & & M11 & 3.61 & -- & 0.32 & $\pm0.03$ \\
 & \vspace{-0.095in} \\
A3560 \dotfill & Cl & NCC & dP07 & 2.02 & $\pm$0.18 & 0.29 & $\pm0.11$ \\
 & \vspace{-0.095in} \\
A3562 \dotfill & Cl & CC & DGM01 & 3.22 & $\pm$0.18 & 0.26 & $\pm0.05$ \\
 & & & M11 & 3.21 & -- & 0.45 & $\pm0.06$ \\
 & \vspace{-0.095in} \\
A3571 \dotfill & Cl & CC & F98 & 4.49 & $\pm$0.11 & 0.23 & $\pm0.03$ \\
 & & & M11 & 4.34 & -- & 0.34 & $\pm0.05$ \\
 & \vspace{-0.095in} \\
A3581 \dotfill & Cl & CC & dP07 & 1.31 & $\pm$0.01 & 0.39 & $\pm0.16$ \\
 & \vspace{-0.095in} \\
A3627 (Norma) \dotfill & Cl & NCC & DGM01 & 4.19 & $\pm$0.12 & 0.28 & $\pm0.02$ \\
 & & & M11 & 3.67 & -- & 0.28 & $\pm0.14$ \\
 & \vspace{-0.095in} \\
A3888 \dotfill & Cl & NCC & dP07 & 6.00 & $\pm$1.04 & 0.24 & $\pm0.13$ \\
 & \vspace{-0.095in} \\
A4038 \dotfill & Cl & CC & M11 & 2.00 & -- & 0.33 & $\pm0.06$ \\
 & \vspace{-0.095in} \\
A4059 \dotfill & Cl & CC & F98 & 2.65 & $\pm$0.08 & 0.38 & $\pm0.07$ \\
 & & & P03 & 3.68 & $\pm$0.18 & 0.33 & $\pm0.05$ \\
 & & & T04 & 2.67 & -- & 0.43 & $\pm0.09$ \\
 & & & dP07 & 2.65 & $\pm$0.12 & 0.43 & $\pm0.07$ \\
 & \vspace{-0.095in} \\
AWM7 \dotfill & Cl & CC & F98 & 2.50 & $\pm$0.06 & 0.33 & $\pm0.11$ \\
 & & & M11 & 2.40 & -- & 0.45 & $\pm0.12$ \\
 & \vspace{-0.095in} \\
Fornax (NGC1399) \dotfill & Gr & CC & F98 & 0.80 & $\pm$0.04 & 0.22 & $\pm0.11$ \\
 & \vspace{-0.095in} \\
HCG42 \dotfill & Gr & CC & RP09 & 0.54 & $\pm$0.04 & 0.15 & $\pm0.06$ \\
 & \vspace{-0.095in} \\
HCG62 \dotfill & Gr & CC & F98 & 0.70 & $\pm$0.03 & 0.17 & $\pm0.07$ \\
 & & & RP09 & 0.67 & $\pm$0.04 & 0.13 & $\pm0.03$ \\
 & & & S14 & 0.67 & $\pm$0.04 & 0.15 & $\pm0.06$ \\
\hline
\end{tabular}
\caption{ \ldots{}continuation of Table \ref{tab:Final_data0}\ldots{}}
\label{tab:Final_data1}
\end{table*}
 
\begin{table*}
\centering
\begin{tabular}{p{1.35in}ccccccc}
\hline \hline
$^{1}$\textbf{Name} & $^{2}$\textbf{Class} & $^{3}$\textbf{Type} & $^{4}$\textbf{Sample} & $^{5}$$\bs{kT_{500}}$ & $^{6}$$\bs{\sigma(kT_{500})}$ & $^{7}$$\bs{\bar{Z}\sub{\textbf{Fe,500}}}$ & $^{8}$$\bs{\sigma(\bar{Z}\sub{\textbf{Fe,500}})}$ \\
\hline
HCG97 (IC5357) \dotfill & Gr & CC & M05 & 0.60 & $\pm$0.03 & 0.16 & $\pm0.02$ \\
 & \vspace{-0.095in} \\
Hydra A (A780) \dotfill & Cl & CC & F98 & 2.38 & $\pm$0.07 & 0.22 & $\pm0.03$ \\
 & & & P03 & 3.68 & $\pm$0.18 & 0.27 & $\pm0.04$ \\
 & & & T04 & 2.27 & -- & 0.29 & $\pm0.04$ \\
 & & & M11 & 2.34 & -- & 0.28 & $\pm0.03$ \\
 & \vspace{-0.095in} \\
IC1459 \dotfill & Gr & NCC & F06 & 0.44 & $\pm$0.03 & 0.07 & $\pm0.03$ \\
 & \vspace{-0.095in} \\
MKW3s \dotfill & Cl & CC & F98 & 2.46 & $\pm$0.06 & 0.27 & $\pm0.05$ \\
 & & & P03 & 2.27 & $\pm$0.18 & 0.31 & $\pm0.05$ \\
 & & & T04 & 2.34 & -- & 0.31 & $\pm0.05$ \\
 & & & dP07 & 2.66 & $\pm$0.05 & 0.34 & $\pm0.05$ \\
 & & & M11 & 2.47 & -- & 0.45 & $\pm0.07$ \\
 & \vspace{-0.095in} \\
MKW4 \dotfill & Gr & CC & F98 & 1.15 & $\pm$0.08 & 0.31 & $\pm0.24$ \\
 & & & RP09 & 1.19 & $\pm$0.08 & 0.24 & $\pm0.07$ \\
 & & & S14 & 1.19 & $\pm$0.08 & 0.51 & $\pm0.39$ \\
 & \vspace{-0.095in} \\
NGC383 \dotfill & Gr & CC & RP09 & 1.11 & $\pm$0.06 & 0.26 & $\pm0.07$ \\
 & \vspace{-0.095in} \\
NGC507 \dotfill & Gr & CC & F98 & 0.84 & $\pm$0.06 & 0.34 & $\pm0.06$ \\
 & & & RP09 & 0.87 & $\pm$0.04 & 0.28 & $\pm0.08$ \\
 & \vspace{-0.095in} \\
NGC533 \dotfill & Gr & CC & P03 & 0.92 & $\pm$0.08 & 0.19 & $\pm0.05$ \\
 & & & T04 & 0.87 & $\pm$0.04 & 0.27 & $\pm0.10$ \\
 & & & RP09 & 0.82 & $\pm$0.05 & 0.17 & $\pm0.05$ \\
 & \vspace{-0.095in} \\
NGC741 (SRGb119) \dotfill & Gr & CC & M05 & 0.91 & $\pm$0.06 & 0.51 & $\pm0.19$ \\
 & & & RP09 & 0.95 & $\pm$0.10 & 0.10 & $\pm0.04$ \\
 & \vspace{-0.095in} \\
NGC1407 \dotfill & Gr & CC & RP09 & 0.68 & $\pm$0.07 & 0.18 & $\pm0.04$ \\
 & \vspace{-0.095in} \\
NGC1550 \dotfill & Gr & CC & S14 & 0.72 & $\pm$0.03 & 0.11 & $\pm0.08$ \\
 & \vspace{-0.095in} \\
NGC2300 \dotfill & Gr & CC & F98 & 0.59 & $\pm$0.03 & 0.17 & $\pm0.13$ \\
 & & & F06 & 0.56 & $\pm$0.03 & 0.33 & $\pm0.20$ \\
 & & & RP09 & 0.52 & $\pm$0.04 & 0.18 & $\pm0.09$ \\
 & \vspace{-0.095in} \\
NGC3411 (SS2b153) \dotfill & Gr & CC & M05 & 0.44 & $\pm$0.02 & 0.15 & $\pm0.06$ \\
 & \vspace{-0.095in} \\
NGC4125 \dotfill & Gr & NCC & RP09 & 0.22 & $\pm$0.08 & 0.15 & $\pm0.09$ \\
 & \vspace{-0.095in} \\
NGC4261 \dotfill & Gr & CC & F06 & 0.83 & $\pm$0.04 & 0.12 & $\pm0.04$ \\
 & \vspace{-0.095in} \\
NGC4325 \dotfill & Gr & CC & RP09 & 0.66 & $\pm$0.03 & 0.18 & $\pm0.06$ \\
 & \vspace{-0.095in} \\
NGC4636 \dotfill & Gr & CC & F06 & 0.57 & $\pm$0.03 & 0.10 & $\pm0.03$ \\
 & \vspace{-0.095in} \\
NGC5044 \dotfill & Gr & CC & F98 & 0.72 & $\pm$0.03 & 0.30 & $\pm0.04$ \\
 & & & F06 & 0.90 & $\pm$0.04 & 0.27 & $\pm0.03$ \\
 & & & RP09 & 0.75 & $\pm$0.04 & 0.21 & $\pm0.05$ \\
 & & & S14 & 0.75 & $\pm$0.04 & 0.12 & $\pm0.02$ \\
 & \vspace{-0.095in} \\
NGC5098 (RGH80) \dotfill & Gr & CC & M05 & 0.70 & $\pm$0.03 & 0.15 & $\pm0.06$ \\
 & \vspace{-0.095in} \\
NGC5129 \dotfill & Gr & CC & M05 & 0.64 & $\pm$0.04 & 0.36 & $\pm0.09$ \\
 & \vspace{-0.095in} \\
NGC5846 \dotfill & Gr & CC & F06 & 0.51 & $\pm$0.02 & 0.14 & $\pm0.03$ \\
 & & & RP09 & 0.44 & $\pm$0.03 & 0.18 & $\pm0.04$ \\
 & \vspace{-0.095in} \\
NGC6338 \dotfill & Cl & CC & RP09 & 1.42 & $\pm$0.13 & 0.18 & $\pm0.08$ \\
 & \vspace{-0.095in} \\
NGC7619 \dotfill & Gr & CC & RP09 & 0.71 & $\pm$0.06 & 0.16 & $\pm0.04$ \\
 & \vspace{-0.095in} \\
NRGb184 (UGC07115) \dotfill & Gr & CC & M05 & 0.84 & $\pm$0.06 & 0.10 & $\pm0.03$ \\
 & \vspace{-0.095in} \\
Ophiuchus \dotfill & Cl & CC & F98 & 6.85 & $\pm$0.21 & 0.19 & $\pm0.05$ \\
 & \vspace{-0.095in} \\
PKS 0745-191 \dotfill & Cl & CC & DGM01 & 5.56 & $\pm$0.17 & 0.24 & $\pm0.04$ \\
 & \vspace{-0.095in} \\
S159-03 (S1101) \dotfill & Cl & CC & P03 & 2.33 & $\pm$0.18 & 0.29 & $\pm0.04$ \\
 & & & T04 & 1.60 & -- & 0.23 & $\pm0.04$ \\
 & & & dP07 & 1.89 & $\pm$0.01 & 0.33 & $\pm0.04$ \\
 & \vspace{-0.095in} \\
Triangulum Austr. \dotfill & Cl & NCC & F98 & 6.71 & $\pm$0.46 & 0.19 & $\pm0.07$ \\
\hline \hline
\end{tabular}
\caption{ \ldots{}completion of Table \ref{tab:Final_data0}. NOTES: \textit{Column 1:} System name (with other common alternative names in parenthesis). \textit{Column 2:} Classification as a group (Gr) or cluster (Cl), as defined in \S \ref{sec:Definitions_and_Derivations}. \textit{Column 3:} Core type as defined in \S \ref{sec:Definitions_and_Derivations}, being either cool-core (CC) or non-cool-core (NCC). \textit{Column 4:} Original sample (Appendix A). \textit{Column 5:} ICM temperature at $r_{500}$, in keV. \textit{Column 6:} Error in $kT_{500}$. We note that errors on the measured mean, emission-weighted ICM temperature are not provided by T04 or M11. Therefore, in these cases, the propagated error on $T_{500}$ is not known. \textit{Column 7:} Mean, mass-weighted iron abundance within $r_{500}$, in $Z_{\textnormal{Fe,\astrosun}}$, assuming the solar abundances of \citet{GS98}. \textit{Column 8:} Error in $\bar{Z}\sub{Fe,500}$.}
\label{tab:Final_data2}
\end{table*}
  
\end{document}